\documentclass[useAMS,usenatbib]{mn2e}
\usepackage{times}
\usepackage{epsfig}
\usepackage{graphicx}
\usepackage{subfigure}
\usepackage{latexsym}
\usepackage{pifont}
\input epsf

\newcommand{\xmm}{{\it XMM~\/}}
\newcommand{\xmmn}{{\it XMM-Newton~\/}}

\newcommand{\chandra}{{\it Chandra~\/}}

\newcommand{\rosat}{{\it ROSAT~\/}}
\newcommand{\einstein}{{\it Einstein~\/}}
\newcommand{\aap}{{\it A\&A~\/}}

\newcommand{\aj}{{\it AJ~\/}}
\newcommand{\apj}{{\it ApJ~\/}}
 
\newcommand{\apjs}{{\it ApJS~\/}}
\newcommand{\araa}{{\it ARA\&A~\/}}
\newcommand{\mnras}{{\it MNRAS~\/}} 
\newcommand{\galex}{{\it GALEX~\/}}

\def\Msun{\hbox{$\rm ~M_{\odot}$}}

\def\ergsec{{\rm ~erg~s^{-1}}}
\def\cm2{{\rm ~cm^{-2}}}

\def\rchi{{$\chi^{2}_{\nu}$}}

\def\ctsec{{\rm ~ct~s^{-1}}}
\def\dg{^{\circ}}
\def\H0{{\rm ~km~s^{-1}~Mpc^{-1}}}

\def\eg{{\it e.g.~\/}}

\def\ie{{\it i.e.~\/}}
\def\cf{{\it cf.~\/}}
\def\la{\mathrel{\hbox{\rlap{\hbox{\lower4pt\hbox{$\sim$}}}{\raise2pt\hbox{$<$}}}}}
\def\ga{\mathrel{\hbox{\rlap{\hbox{\lower4pt\hbox{$\sim$}}}{\raise2pt\hbox{$>$}}}}}
\def\d25{D$_{25}$}
\def\nh{{$N_{H}$}}

\def\ovii{O{\small VII}$~$}
\def\hii{H{\small II}$~$}
\def\lx{L$_{\rm X}$}

\def\deg{\hbox{$^\circ$~\/}}
\def\arcm{\hbox{$^\prime$}}
\def\arcs{\hbox{$^{\prime\prime}$}}

\def\eps@scaling{1.0}%
\newcommand\plottwo[2]{%
  \centering
  \leavevmode
  \columnwidth=.45\columnwidth
  \includegraphics[width={\eps@scaling\columnwidth}]{#1}%
  \hfil
  \includegraphics[width={\eps@scaling\columnwidth}]{#2}%
}%

\title[Soft X-ray emission from the inner disk of M33]{Soft X-ray emission from 
the inner disk of M33}
\author[R.A.Owen  \& R.S. Warwick]
	{R.A.\ Owen, R.S.\ Warwick \\
X-ray \& Observational Astronomy Group, Dept. of Physics \& Astronomy, 
University of Leicester, University Road, Leicester LE1 7RH, U.K.\\}

\pagerange{\pageref{firstpage}--\pageref{lastpage}}

\pubyear{2009}

\begin{document}

\maketitle

\label{firstpage}


\begin{abstract}


We present a study, based on  archival \xmmn observations, of the extended X-ray
emission associated  with the inner  disk of M33.  After the exclusion  of point
sources with \lx  $> 2 \times 10^{35} \ergsec$ (0.3-6  keV), we investigate both
the  morphology and  spectrum of  the  residual X-ray  emission, comprising  the
integrated signal from unresolved  discrete sources and diffuse components. This
residual emission  has a  soft X-ray  spectrum which  can be fitted  with a
two-temperature thermal model,  with $\rm kT \approx 0.2$ keV  and $\approx 0.6$ keV,
the  cooler  component accounting  for  the bulk  of  the  luminosity. There  is
evidence that the X-ray emitting plasma has a subsolar metal abundance.
The soft  X-ray surface brightness distribution shows a strong  correlation with FUV emission
and since the latter serves as a tracer of the inner spiral arms of M33, this is
indicative of a close connection  between recent star-formation activity and the
production of soft X-rays. Within 3.5 kpc  of the nucleus of M33, the soft X-ray
and   FUV  surface   brightness  distributions   exhibit  similar  radial
profiles.  The  implication is  that  the ratio  of  the  soft X-ray  luminosity
(0.3-2.0 keV) per unit disk area to  the star formation rate (SFR) per unit disk
area remains  fairly constant within this  inner disk region. We  derive a value
for   this  ratio   of  $1-1.5   \times  10^{39}   \ergsec(\Msun  {\rm
yr}^{-1})^{-1}$, which is towards the top  of the range of similar estimates for
several other nearby face-on spiral galaxies (\eg M51, M83).   In the same region, the ratio of
soft  X-ray  luminosity  to  stellar  mass  (the  latter  derived  from  K-band
photometry) is  $4 \times 10^{28} \ergsec  \Msun^{-1}$, a factor  of 5-10 higher
than is  typical of  dwarf elliptical galaxies (\eg M32, NGC3379),  suggesting that 
10-20\%  of the
unresolved emission seen in M33 may originate in its old stellar population.
The remainder of the soft X-ray emission is found to be equally split between 
two spatial components, one which closely traces the spiral arms of the galaxy 
and the other more smoothly distributed across the inner disk of M33. The former
must represent a highly clumped low-filling factor component linked to sites of
recent or ongoing star formation, encompassing \hii regions, X-ray source complexes
and X-ray superbubbles, whereas the distribution of the latter gives few clues as 
to its exact origin.

\end{abstract}

\begin{keywords}

Galaxies: ISM -- galaxies: spiral -- X-rays: galaxies

\end{keywords}


\section{Introduction}
\label{sec:m33intro}

On the basis of current observations,  the X-ray emission of spiral galaxies may
be separated into several components.  A very significant contribution generally
arises  from a  set  of resolved  point  sources which  correspond  to the  most
luminous  examples of the  galactic X-ray  binary population,  encompassing both
high-mass  (HMXB)  and low-mass  (LMXB)  systems.  To  this  must  be added  the
integrated emission  from large  numbers of lower-luminosity  sources, including
supernova  remnants (SNRs), more  quiescent forms  of X-ray  binary, cataclysmic
variables, and coronally  active stars.  Finally we need  to include the thermal
X-ray  signal  emanating from  concentrations  of  truly  diffuse million-K  gas
resident both in  the galactic disk and potentially  extending into the galactic
halo.  The extent to which these components can be distinguished from each other
depends on the orientation and distance  of the target galaxy and, of course, on
the spatial resolution and threshold sensitivity afforded by the observation.

Early X-ray  studies based on  \einstein observations focussed on  the brightest
point  sources  seen  in  a  number  of nearby  galaxies,  whilst  also  finding
underlying extended  emission in complex  structures (\citealt{fabbiano89}). The
superior spatial resolution and soft response of {\it Rosat} further progressed
the field, for example allowing the extent of the apparently diffuse emission to
be traced  both in  and out of  the plane  of the disk  for face-on  and edge-on
systems  respectively (\eg  \citealt{read97}; \citealt{dahlem98}).   More recent
investigations, which  exploit the enhanced collecting  area, spatial resolution
and  spectral  sensitivity   of  \xmmn  and  {\it  Chandra},   have  probed  the
point-source populations in galaxies  to much fainter thresholds than previously
attainable        (\eg       \citealt{strickland04a};       \citealt{colbert04}).  
The X-ray luminosity functions (XLF) typical of the HMXB
and   LMXB   populations   seen   in   spiral   galaxies   have   thereby   been
determined (\citealt{grimm05}; \citealt{kilgard05}; \citealt{fabbiano06}). 
 By extrapolation of the  XLF to fluxes below the threshold at
which individual sources  can be resolved, the total  aggregated luminosity of a
particular class of source may also be inferred (\citealt{sazonov06})

In  two previous papers  (\citealt{warwick07}, hereafter  W07; \citealt{owen09},
hereafter OW09),  we studied  the morphology of  the underlying  unresolved soft
X-ray emission  in a  sample of six  nearby face-on  spiral galaxies and  made a
comparison with other wavelength, particularly FUV, measurements.  These studies
confirmed the close connection  between recent star-formation and the production
of  soft  X-rays.   We  also  investigated  how the  soft  X-ray  luminosity  to
star-formation rate  (SFR) ratio varied radially within  the individual galaxies
as well as  from galaxy  to  galaxy,  and found  some  evidence  for an  enhanced
efficiency in the production of soft  X-rays in regions of high SFR density. The
link between soft X-rays and star  formation can be further explored by studying
galaxies closer to us,  in which the majority of HMXBs and  LMXBs can be readily
resolved, so as to provide a relatively clear perspective of the underlying soft
X-ray emission. It is in this context that we focus in the present paper on M33.

M33   is   an  Sc   spiral   galaxy   with   an  inclination   of   $56^{\circ}$
(\citealt{zaritsky89}) at  a distance of  795 kpc (\citealt{vdburgh91}).  As the
the  third largest  galaxy in  the Local  Group, the  relative proximity  of M33
permits detailed study  of its discrete X-ray source  population and enables the
separation  of  these sources  from  any  residual emission  in  the  disk to  a
relatively  faint threshold.  The relatively  low Galactic  foreground  \nh~ ($7.5
\times 10^{20}  \cm2$; \citealt{kalberla05})  in the direction  of M33, the  low to
moderate inclination of the galaxy, and its relatively high SFR in comparison to
other nearby systems (0.3-0.7\Msun yr$^{-1}$,  \citealt{hippelein03}) make M33
an ideal candidate for the present study.

Early  \einstein observations  of  M33 revealed  a  total of  17 point  sources,
including  the  bright   ULX  M33  X-8  (\citealt{long81};  \citealt{markert83};
\citealt{trinchieri88}).  \citet{trinchieri88}  also   found  evidence  for  the
presence of a  soft diffuse component in the plane of  the galactic disk. \rosat
observations   (\citealt{schulman95};  \citealt{long96})   expanded   the  known
population of  X-ray point sources  in the  direction of M33  to a total  of 184
(\citealt{haberl01}). More recently, using \xmmn  data, a total of 350-400 X-ray
point  sources to a  limiting X-ray  luminosity of  $10^{35} \ergsec$  have been
identified    and   categorized    (\citealt{foschini04};   \citealt{pietsch04};
\citealt{grimm05};  \citealt{misanovic06};  \citealt{grimm07}).   These  results
have   been   complemented   by   recent   studies   of   M33   using   \chandra
(\citealt{plucinsky08};  \citealt{williams08}) which  have increased  the source
statistics particularly in the confused inner regions of the galaxy. Most of the
emphasis  to  date has  been  on  the X-ray  emission  from  point sources,  but
\citet{haberl01},  \citet{pietsch04}   and  \citet{tullman08}  have   noted  the
presence  of seemingly  diffuse structures  along the  spiral arms.  It  is this
latter component which we examine in this paper.

Here, we  focus on the spectral  and spatial properties of  the unresolved X-ray
emission   from  the   inner  disk   of   M33,  deduced   from  archival   \xmmn
observations. In \S\ref{sec:m33obs}, we describe the set of observations used to
construct soft-band X-ray images and  outline the methods used in the subsequent
data analysis.  In \S\ref{sec:m33lx}, we  briefly examine the properties  of the
resolved discrete  point source population  and the spatial distribution  of the
residual  soft X-ray  emission, which  remains after  the contribution  of these
sources is excluded.  We go on  to compare the soft X-ray morphology with \galex
FUV  measurements and 2MASS K band data. We follow  this with  spectral analysis  
of the  bright point source population and unresolved    residual       emission
(\S\ref{sec:m33spectrum}). We examine the  spatial extent of the X-ray emission
in  comparison  with star  formation  data and the underlying  mass distribution 
(\S\ref{sec:m33star}) for M33, and compare these relationships to those found in 
other galaxies.  
Finally   we  discuss   the  implication   of   our  results
(\S\ref{sec:m33disc})        and         summarize        our        conclusions
(\S\ref{sec:m33conc}). 


\begin{table*}
\caption{Details of the \xmmn observations of M33.}
\small
 \centering
  \begin{tabular}{lcccccccccc}
\hline
Galaxy    & Observation ID & Start Date  & Filter~$^{a}$   & & \multicolumn{2}{c}{Target co-ordinates}  & & \multicolumn{2}{c}{Useful exposure (ks)} \\
      &                & (yyyy-mm-dd) & pn/MOS1/MOS2    & & RA (J2000)       & Dec (J2000)             & & pn            & MOS 1+2
    \\
\hline
M33  &  0102640101$^{c}$  &  2000-08-04  &  M/-/-   & & $01^h33^m51.0^s$ & $+30\deg39\arcm37\arcs$~$^{b}$ & & 7.1  &  -     \\
   & 0102640201  &  2000-08-04   &  M/M/M  & &  $01^h34^m40.0^s$ & $+30\deg57\arcm48\arcs$   & & 11.8   & 31.5   \\
  &  0102640301  &  2000-08-07  &  M/M/Tn  & &  $01^h33^m32.0^s$ & $+30\deg52\arcm13\arcs$   & &  3.6   & 9.5    \\
  &  0102640401  &  2000-08-02  &  Tk/Tk/Tk  & & $01^h32^m51.0^s$ & $+30\deg36\arcm49\arcs$  & &  9.1  &  23.1  \\
  &  0102640501  &  2001-07-05  &  M/M/M    & &  $01^h33^m02.0^s$ & $+30\deg21\arcm24\arcs$  & &  9.2  &  23.1  \\
  &  0102640601  &  2001-07-05  &  M/M/M    & &  $01^h34^m08.0^s$ & $+30\deg46\arcm06\arcs$  & &  4.5  &  11.9  \\
  &  0102640701  &  2001-07-05  &  M/M/M    & &  $01^h34^m10.0^s$ & $+30\deg27\arcm00\arcs$  & &  6.9  &  21.9  \\
  &  0102640801  &  2001-07-07  &  -/M/M    & &  $01^h34^m51.0^s$ & $+30\deg42\arcm22\arcs$  & &  -  &  3.2  \\
  &  0102640901  &  2001-07-08  &  M/M/M    & &  $01^h34^m04.0^s$ & $+30\deg57\arcm25\arcs$  & &  3.9  &  11.2  \\
  &  0102641001  &  2001-07-08  &  M/M/M    & &  $01^h33^m07.0^s$ & $+30\deg45\arcm02\arcs$  & &  1.5  &  16.3  \\
  &  0102641101  &  2001-07-08  &  M/M/M    & &  $01^h32^m46.0^s$ & $+30\deg28\arcm19\arcs$  & &  8.0  &  21.0  \\
  &  0102641201  &  2000-08-02  &  Tk/Tk/Tk  & & $01^h33^m38.0^s$ & $+30\deg21\arcm49\arcs$  & &  12.0  &  7.2  \\
  &  0102642001  &  2001-08-15  &  M/M/M     & & $01^h34^m51.0^s$ & $+30\deg42\arcm22\arcs$  & &  8.8  &  22.3  \\
  &  0102642101  &  2002-01-25  &  M/M/M     & & $01^h34^m34.0^s$ & $+30\deg34\arcm11\arcs$  & &  10.0  &  24.3  \\
  &  0102642201  &  2002-01-25  &  M/M/M     & & $01^h34^m56.0^s$ & $+30\deg50\arcm52\arcs$  & &  11.5  &  27.3  \\
  &  0102642301$^{c}$  &  2002-01-27  &  M/M/M     & & $01^h33^m33.0^s$ & $+30\deg33\arcm07\arcs$  & &  9.9  &  24.1  \\
  &  0141980101  &  2003-07-11  &  M/M/M     & & $01^h33^m07.0^s$ & $+30\deg45\arcm02\arcs$  & &  6.2  &  13.2  \\
  &  0141980301  &  2003-07-25  &  -/M/M     & & $01^h34^m08.0^s$ & $+30\deg46\arcm06\arcs$  & &  -   &  1.2  \\
  &  0141980501  &  2003-01-22  &  M/M/M     & & $01^h33^m51.0^s$ & $+30\deg39\arcm37\arcs$  & &  1.9  &  17.0  \\
  &  0141980601  &  2003-01-23  &  M/M/M     & & $01^h32^m51.0^s$ & $+30\deg36\arcm49\arcs$  & &  11.0  &  25.9  \\
  &  0141980701  &  2003-01-24  &  M/M/M     & & $01^h33^m38.0^s$ & $+30\deg21\arcm49\arcs$  & &  4.4  &  11.4  \\
  &  0141980801$^{c}$  &  2003-02-12  &  M/M/M     & & $01^h33^m51.0^s$ & $+30\deg39\arcm37\arcs$  & &  7.8  &  19.8  \\


\hline
\end{tabular}
\\
\raggedright
$^{a}$ - Tn = thin filter, M = medium filter, Tk = thick filter \\
$^{b}$ - Assumed position of the galactic nucleus.\\
$^{c}$ - Observations used for spectral analysis.\\
\label{table:m33obs}
\end{table*}



\begin{figure*}
\centering
\rotatebox{270}{\scalebox{0.45}{\includegraphics{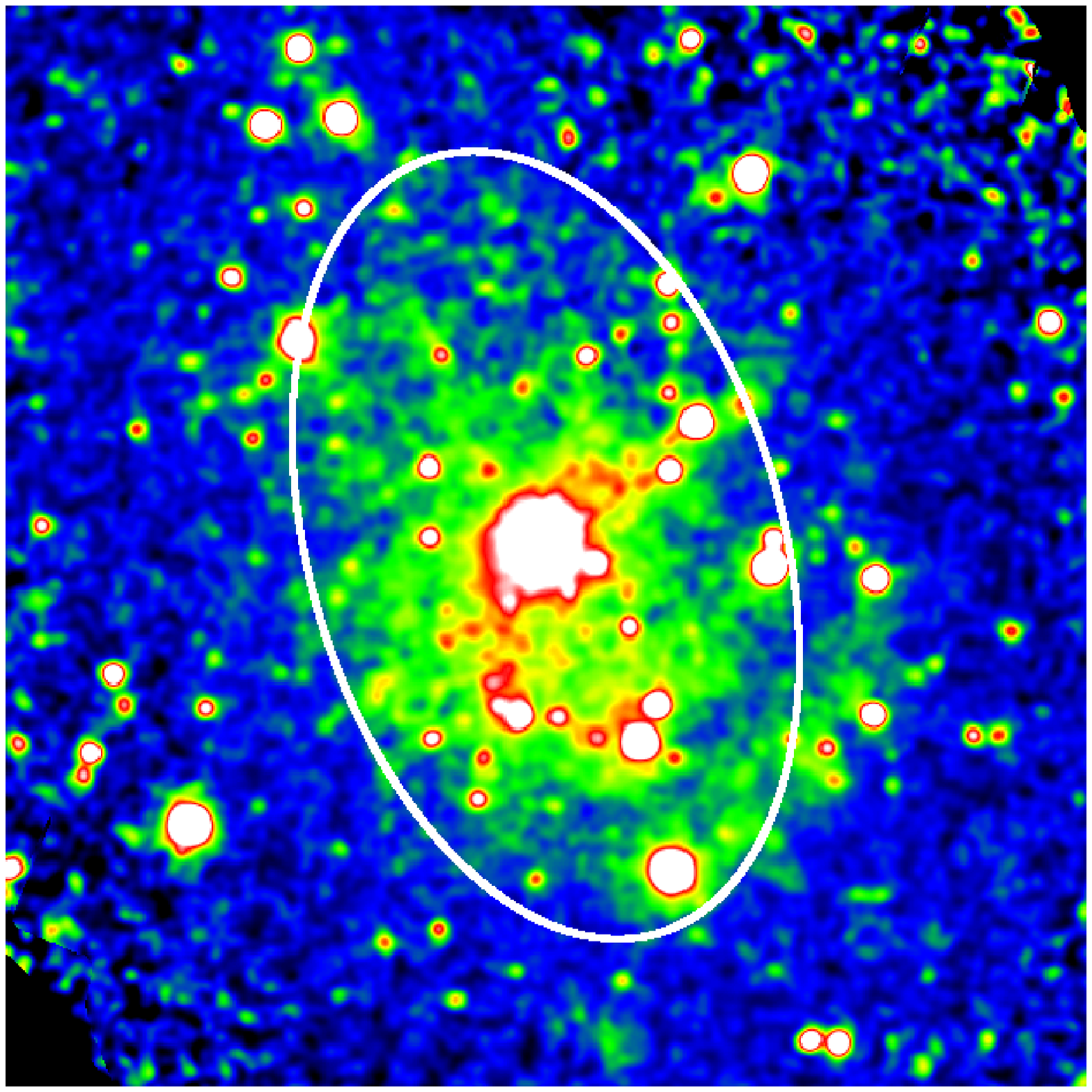}}}
\rotatebox{270}{\scalebox{0.45}{\includegraphics{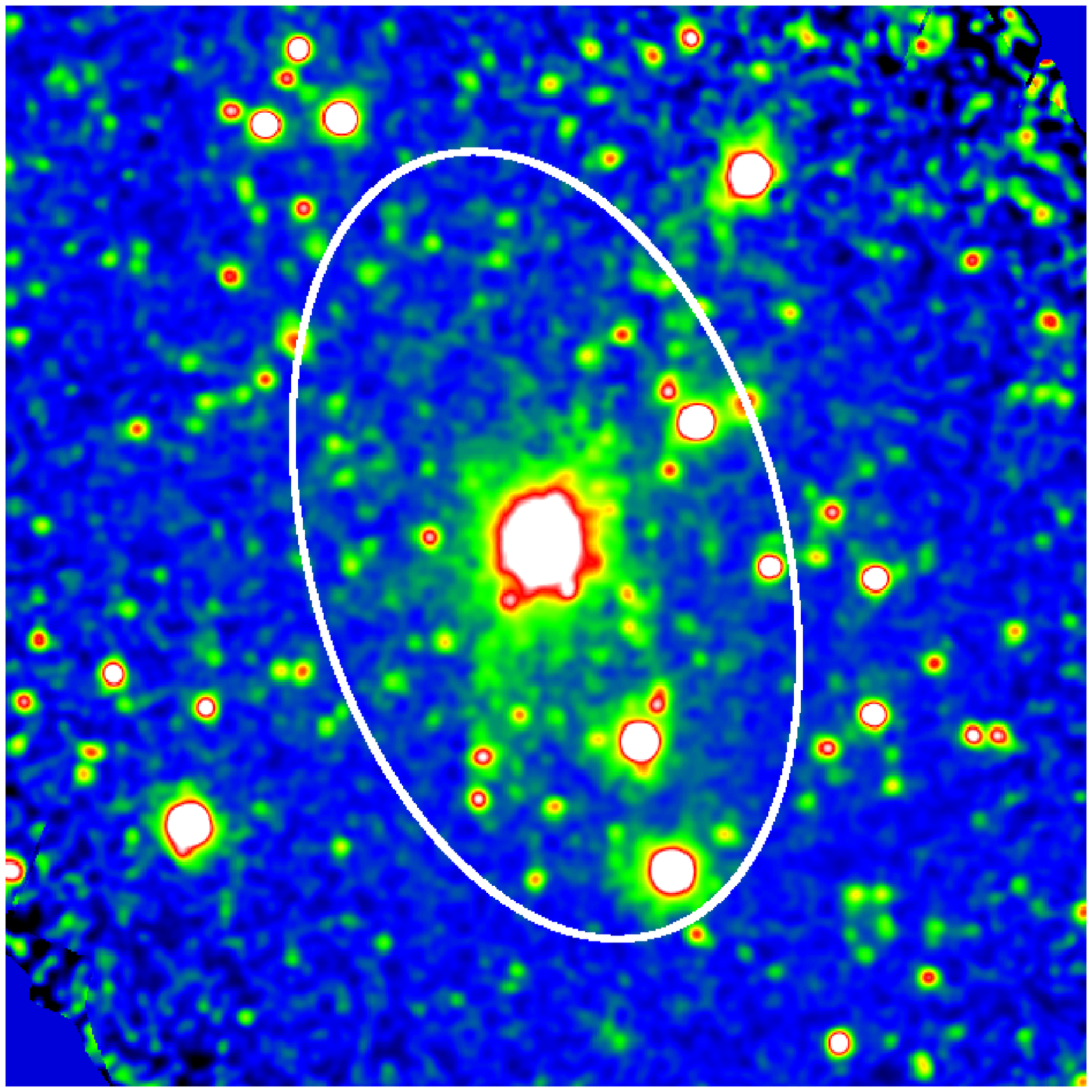}}}
\rotatebox{270}{\scalebox{0.45}{\includegraphics{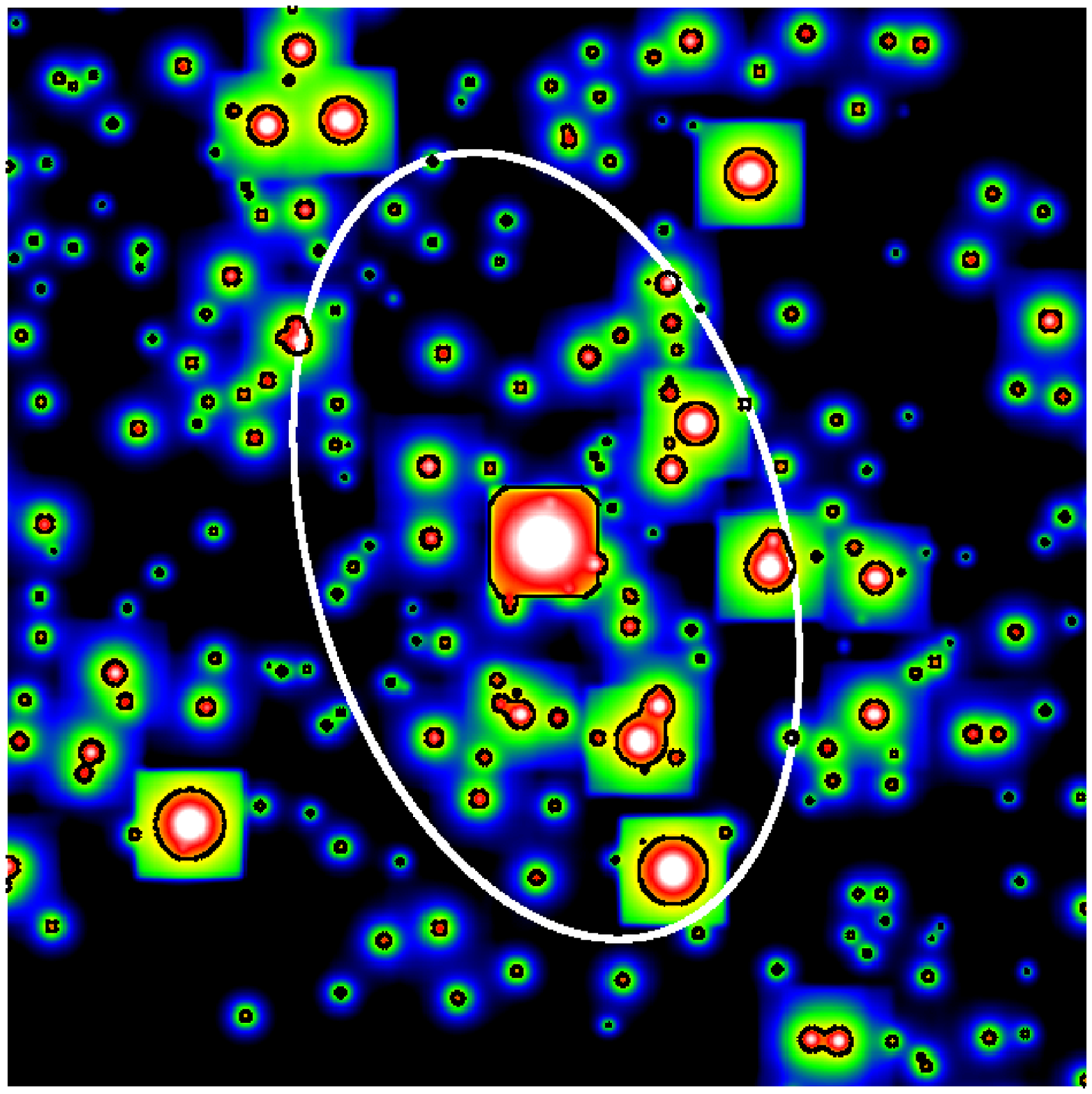}}}
\rotatebox{270}{\scalebox{0.45}{\includegraphics{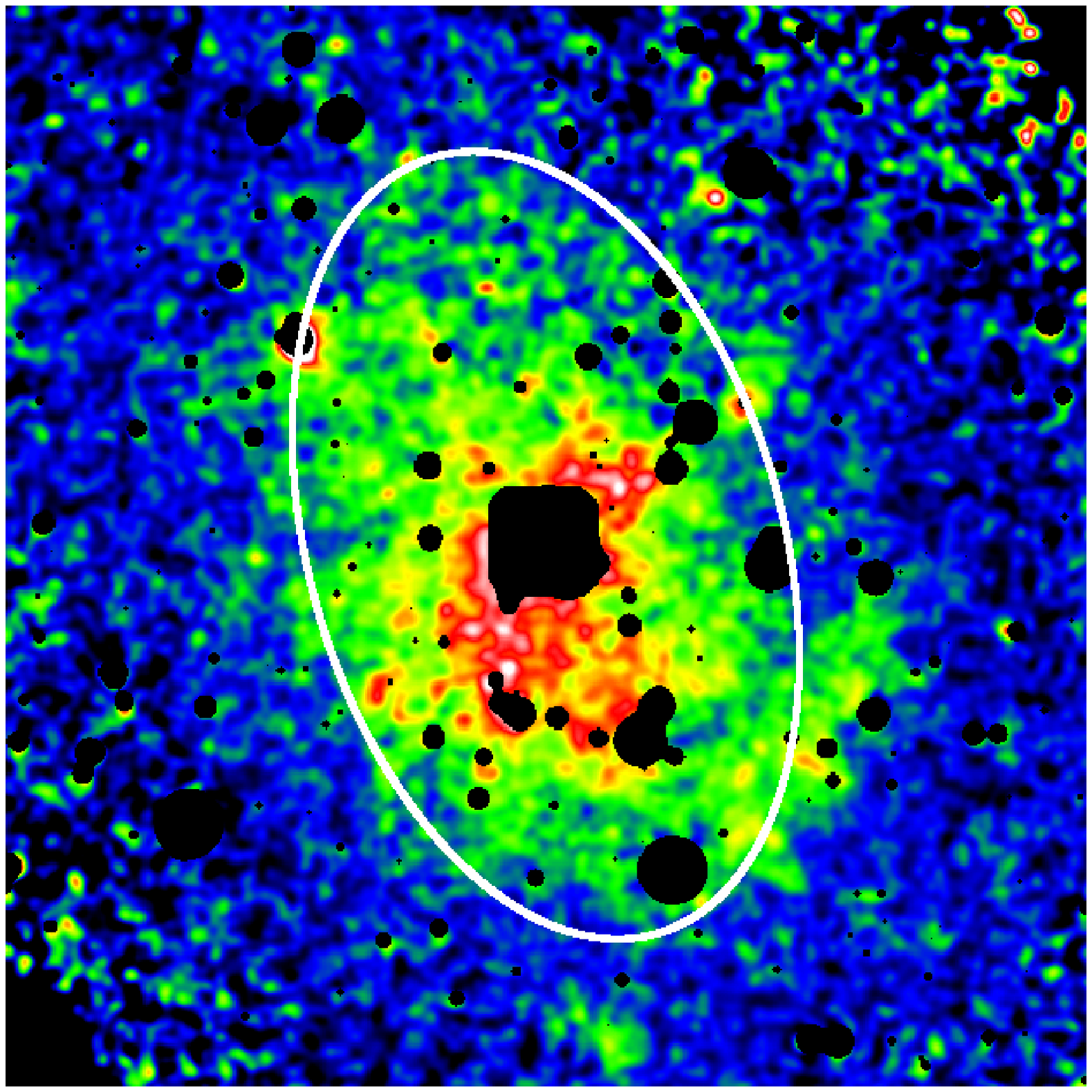}}}
\rotatebox{270}{\scalebox{0.45}{\includegraphics{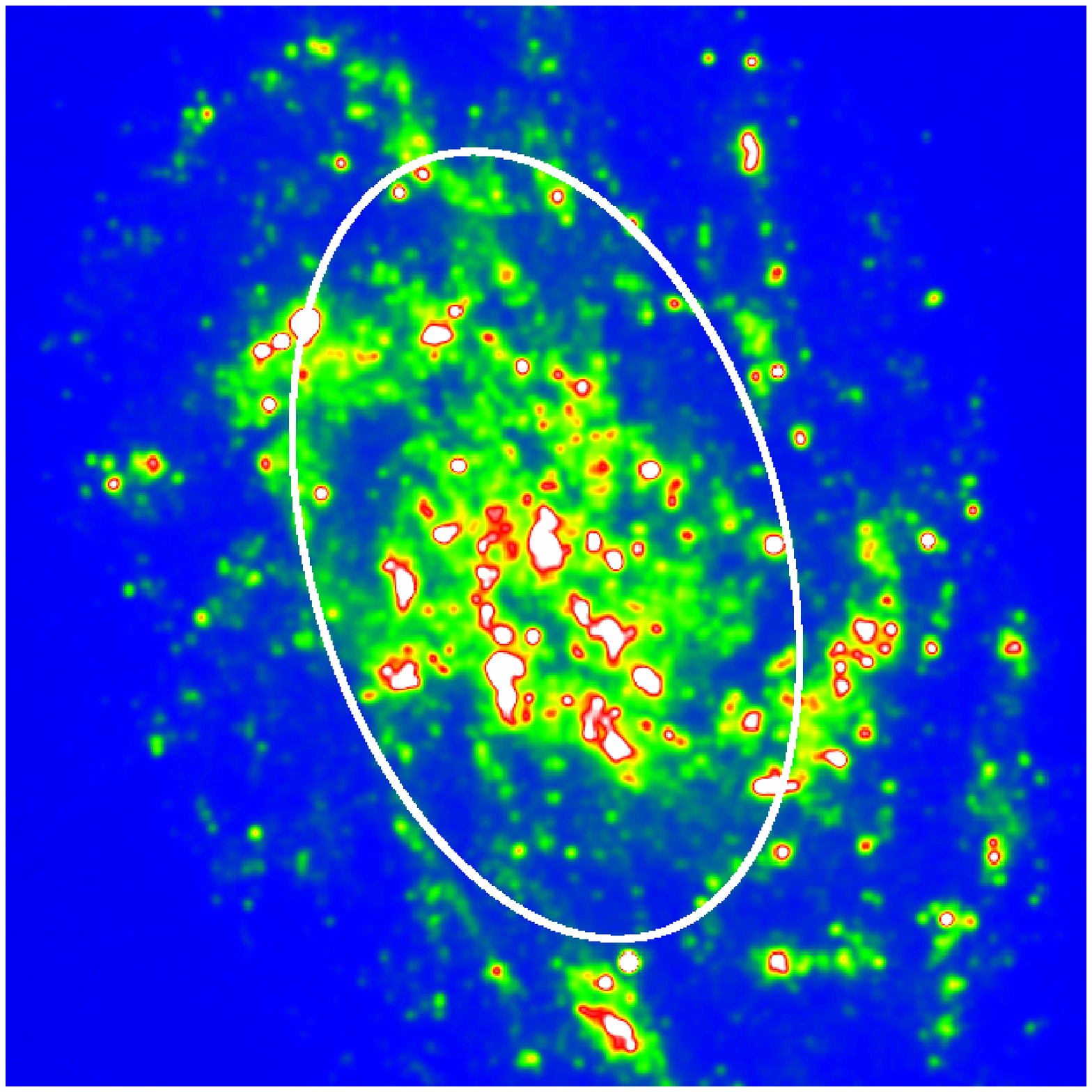}}}
\rotatebox{270}{\scalebox{0.45}{\includegraphics{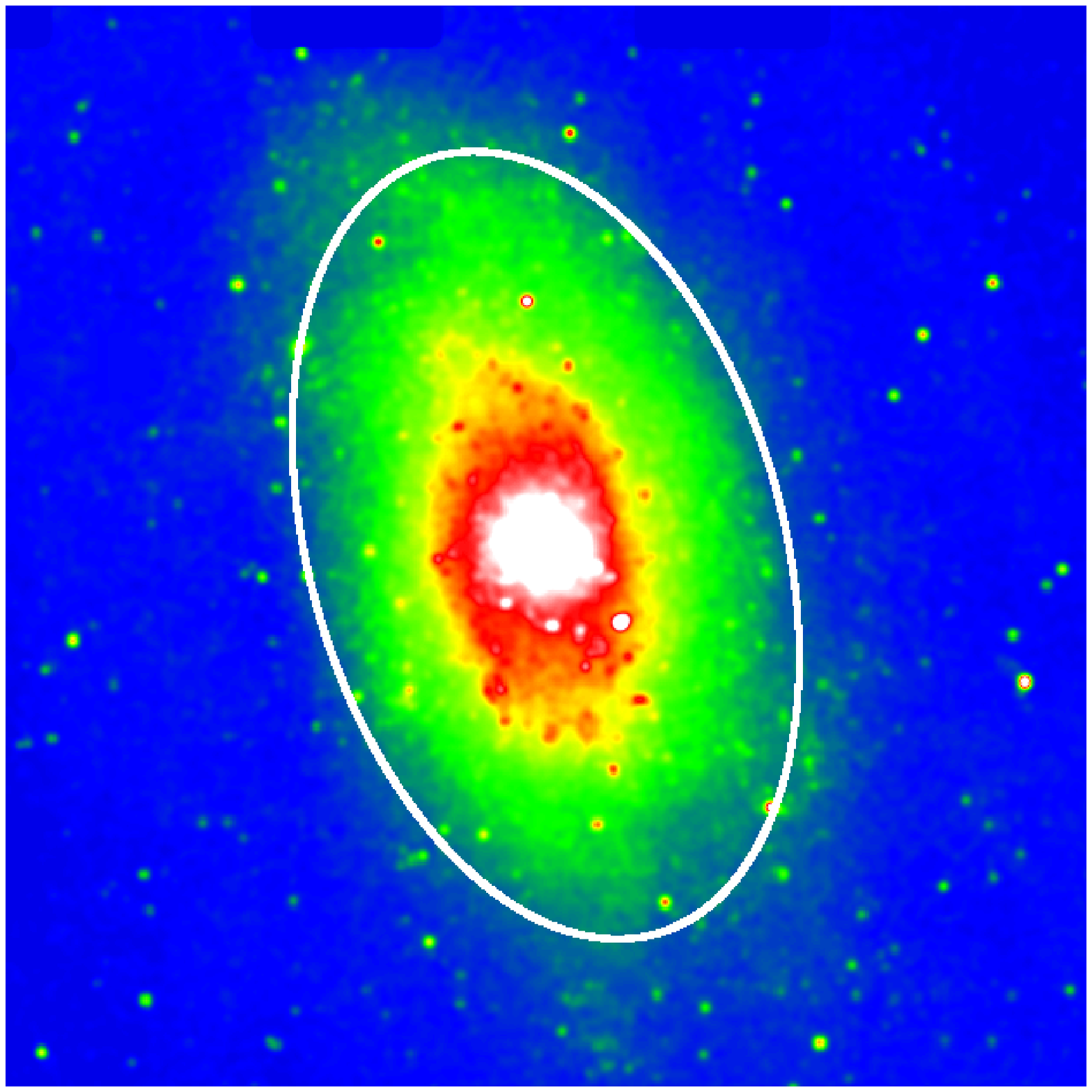}}}
\caption{{\it Top-left panel:}  Adaptively smoothed version of
the \xmmn pn+MOS image of M33 in the soft (0.3-1 keV) band. 
{\it Top-right panel:} The same image in the medium (1--2 keV) band. 
{\it Middle-left panel:} Simulated image of the bright sources in M33, 
with the source mask contours overlaid. The bright central region is 
dominated by the ULX M33 X-8. {\it Middle-right panel:} 
The residual emission in M33 in the 0.3-1 keV band obtained
by subtracting the bright source model and applying the spatial mask. 
{\it Bottom-left  panel:} The {\it GALEX} FUV  ($\lambda_{eff}\approx1528$\AA) image on
the same spatial scale as the X-ray  data. {\it Bottom-right panel:} 2MASS K band
image on the same spatial scale as  the X-ray data.  
The ellipse shown in each image represents the area of the galaxy used for 
X-ray analysis (see text for details).
All the images are displayed with logarithmic amplitude scaling and are
40\arcm on a side.
}
\label{fig:m33xuv1}
\end{figure*}


\section{Observations and Data Reduction}
\label{sec:m33obs}

In two previous papers (W07, OW09) we  reported the results of an \xmmn study of
six  nearby   face-on  galaxies,  namely   M101,  M83,  M51,  M74,   NGC300  and
NGC3184. Here we  use similar methods to incorporate M33  into our sample. Since
the \d25 diameter of M33 is 70\arcm, which is more than a factor two larger than
the  30\arcmin~field of  view  of the  European  Photon  Imaging Camera  (EPIC)
instrument  on {\it  XMM-Newton},  multiple-pointings are  needed  to give  good
coverage of  the inner disk of  M33.  The survey  conducted by \citet{pietsch04}
matches this requirement  and in the present work we  have accessed the revelant
datasets  via the  \xmmn  public  archive.  Details  of  the observations  which
comprise the M33 survey are summarised in Table~\ref{table:m33obs}.

Data reduction was based on SAS  v8.0. The datasets were screened for periods of
high background through the creation  of full-field 10--15 keV lightcurves.  MOS
data were excluded  when the 10--15 keV  count-rate in a 100 s  bin exceeded 0.2
$\ctsec$, whereas pn data were excluded when the count rate in the same waveband
exceeded  2  $\ctsec$.   The   resulting  exposure  times  after  filtering  the
observations, as reported  in Table~\ref{table:m33obs}, range from 1.5  to 13 ks
in the pn camera, with typically somewhat greater exposure in the individual MOS
cameras.   Several of  the pn  datasets which  include the  bright ULX  near the
nucleus of  M33 are affected  by Out  Of Time (OoT)  events. The SAS  task
{\it epchain}  was  used to  create  ``OoT  event lists'',  which
simulate the spatial distribution of  OoT events in the observations. From these
event  lists, images of  the distribution  of OoT  events were  produced.  These
images  were subtracted from  the raw  images produced  from the  original event
lists to  produce images  cleaned of OoT  events, which  could then be  used for
imaging analysis. Additionally, the target observations were examined to look for 
evidence of solar wind charge exchange (SWCX) emission, using the methods 
detailed in \citet{carter08}. One observation was found to be highly contaminated 
(obsid:0141980201) and was thus excluded from the dataset.

In OW09 it was shown how an appropriate spatial mask could be used to excise the
bright source  population from the  residual emission, thereby  greatly reducing
the  ``contamination'' due to  the former.  Here we  employ a  slightly modified
version of this  image processing technique, necessitated by  the need to mosaic
the data sets from the M33 survey observations.
 

\subsection{Image Construction}
\label{sec:m33images}

The methods we have employed to produce the X-ray images of M33 mirror the
procedures described in OW09. For consistency,  we use the same soft, medium and
hard energy bands (0.3--1 keV, 1--2  keV and 2--6 keV respectively), although in
the  present paper  emphasis is  placed  on the  soft band,  which contains  the
majority of the diffuse signal.  Images and exposure maps were extracted for all
three  bands  and cameras  for  each  observation.   A constant  particle  rate,
estimated  from  the  corners of  the  detector  not  exposed  to the  sky,  was
subtracted  from each image  and the  set of  images for  a given  bandpass were
co-added using the method detailed in  OW09, with a position offset dependent on
the pointing direction of the observation.  As the majority of observations were
conducted using the medium filter,  factors were derived to convert observations
with  the thin  and thick  filters to  the same  relative intensity  scale.  The
exposure maps  were likewise mosaiced to  produce a single exposure  map for the
extended field.   Finally, flat-fielding was  achieved by dividing  the mosaiced
image for each band by the relevant exposure map.

At  this stage  it proved  necessary  to apply  an iterative  adjustment to  the
particle background estimates, so as to  create a visually flat image.  This was
achieved  by  imposing  a  mask  on  the flat-fielded  images  consisting  of  a
point-source mask (see \S\ref{sec:m33imagemask}) coupled with an elliptical mask
extending to a major axis radius of 15\arcm~from the nucleus of the galaxy.  The
ellipticity of  the latter was fixed  at the major/minor axis  ratio reported in
the RC3 catalogue  (\citealt{devaucouleurs91}) for M33, that is  1.7:1, with the
position angle of the  major axis set to 22\deg (\citealt{corbelli07}). We
next imposed the  requirement that the level of the  X-ray sky background should
be constant  in the  flat-fielded images  outside of the  masked region  (\ie we
assume that there  are no significant gradients in  soft X-ray background across
the full field of the survey and also that there is negligible contribution from
M33 in  this outer region).  This  did in fact  appear to be justified  based on
visual  inspection of  the  raw images.   Next,  the average  level  of the  sky
background  in the un-masked  region of  the image  was calculated.   Using this
average  value, the  particle background  rates for  each observation  then were
revised, so as to force the sky  background outside of the masked region {\it in
each  observation} to  this average.   The resulting  mosaiced  and flat-fielded
images of M33 in the soft and medium bands are shown in Fig.  \ref{fig:m33xuv1}.

We  note  that  given the  above,  radial  analysis  of the  surface  brightness
distributions contained within  the M33 images is only  justified for the region
extending from the  nucleus to a major-axis radius of  15\arcm.  This equates to
$42\%$  of the \d25  region and  to a  radial linear  extent of  3.5 kpc  at the
distance of M33. We describe this region hereafter as the inner disk of M33.


\subsection{Spatial masking of bright sources}
\label{sec:m33imagemask}

The catalogue  of sources used to  produce a bright-source spatial  mask for M33
was taken from  the work of \citet{misanovic06}, who  reported 350 X-ray sources
across the \d25 disk of M33 above a luminosity of $2 \times10^{35}\ergsec$. 
Ninety-two of  these sources  lie within  the  elliptical region
defined  in  \S\ref{sec:m33images}  (see  Table  \ref{table:m33gal_sources}  for
summary   details).   Following  the   methods  described   in  OW09,   a  model
``bright-source  image'' was  created for  M33.   A surface  brightness cut  was
applied   to   this    image   at   a   level   of    0.07   pn+MOS1+MOS2   $\rm
ct~ks^{-1}~pixel^{-1}$ to produce the  source mask (see Fig.\ref{fig:m33xuv1}).
We estimate the ``spillover''
fraction  of  this  mask, \ie  the  fraction  of  the bright-source  signal  not
contained within  the masked region,  to be $\approx 4\%$.   This mask was  used to
divide  the inner  disk region  into  two components,  namely a  ``bright-source
region''  and a ``residual-emission  region'' (using  the terminology  of OW09),
which could  then be subject to  appropriate spatial and  spectral 
analysis\footnote{The source mask is square in the region of the very bright central 
source due to clipping of the specified surface brightness contour at the edge of 
the $4\arcm\times4\arcm$ sub-image used to represent the point spread function. 
The leakage of flux from the bright central source is calculated with respect to 
the mask actually employed.}.  
In the image  analysis, low-level contamination of the  residual-emission signal by
the spillover from the bright  sources was further suppressed by subtracting the
simulated image  from the corresponding  pn+MOS image and reimposing  the source
mask. In  contrast, for  spectral analysis, the  spillover was accounted  for by
including a  bright-source contribution in the spectral  fit of the
residual emission.


\begin{table*}
\caption{The parameters of the region investigated in this paper. 
}
\centering
\begin{tabular}{lccccccc}
\hline
Galaxy  &  X-ray extent$^{a}$ & Major-axis PA & Threshold~\lx$^{b}$ & Number in 
& \multicolumn{1}{c}{Number of high \lx~sources} \\
   & (\arcm)     & (\deg) &  ($10^{35}\ergsec$) &  Source List
&  (\lx $> 5 \times 10^{38}\ergsec$) \\
\hline
M33   &  30.0/17.6  & 22 &  2.0 & 92  & 1  \\
\hline    
\end{tabular}
\\
$^{a}$ - The major/minor axis diameters of the ``X-ray extraction region''. \\
$^{b}$ - Nominal \lx~threshold applied in the 0.3--6 keV band 
in defining the bright-source sample. \\
\label{table:m33gal_sources}
\end{table*}



\begin{table*}
\caption{Breakdown of the total X-ray luminosity of M33.}
\centering
\begin{tabular}{lcccccc}
\hline

Galaxy &  Spillover/Area  & Component   & \multicolumn{4}{c}{\lx ($10^{38}\ergsec$)}
     \\
& Factors (\%) & & (0.3-1 keV)        & (1-2 keV) & (2-6 keV) & (0.3-6 keV)   \\
\hline
M33  & 4/6 & Bright sources$^{a}$     &  5.6      &  4.0    &  6.2   &  15.8 \\
     &     & Unresolved sources$^{b}$ & [0.06]    & [0.04]  & [0.07] \\
     &     & Residual emission        &  1.1      &  0.1    &  -     &  1.2  \\
     &     & Total measured           &  6.7      &  4.1    &  6.2   &  17.0  \\
\\

\hline    \\
\end{tabular} 
\\
$^{a}$ - Down to a threshold \lx = $2 \times 10^{35}\ergsec$. \\
$^{b}$ - Extrapolated to the lower limit \lx = $1 \times 10^{34}\ergsec$ \\
\label{table:m33gal_xray}
\end{table*}



\subsection{Spectral Extraction}
\label{sec:m33diff_extr}

The soft-band  image in Fig.~\ref{fig:m33xuv1} demonstrates the  existence of an
extended  emission component  in  addition  to the  population  of bright  point
sources.  On the  basis  of the  approach  described earlier,  we extracted  the
integrated pn spectrum of both  the bright-source region (bounded by the spatial
mask)  and  the  residual-emission  region  (corresponding  to  the  full  X-ray
extraction region  less the source-masked area).  This process was carried  out for the
three observations  indicated in Table \ref{table:m33obs}, which  were chosen as
they  encompass the  central region  of the  galaxy with  reasonable observation
times.  A narrow strip of one CCD  contaminated by Out of Time (OoT) events from
the ULX  was excised in each observation.   The SAS tools {\it  arfgen} and {\it
rmfgen}  were used  to produce  appropriate  Auxiliary Response  File (ARF)  and
Response Matrix File (RMF) files for the source and residual galaxy regions, and
the counts  recorded in adjacent (raw)  spectral channels were summed  to give a
minimum of 20 counts per spectral bin in the final set of spectra.

The large  extent of the  bright-source and residual-emission regions  makes the
process of determining appropriate background spectra more complex than for most
\xmmn  applications. In  OW09,  we  used appropriately  scaled  spectra from  an
annulus  surrounding the defined  galaxy region  and the  corner regions  of the
detector to approximate  the background.  This process is not  viable for M33 as
the central  galaxy field covers too  large a fraction  of the EPIC pn  field of
view, thus  making the extraction  of an adequate background  region impossible.
We therefore used a combination  of ``blank-sky'' fields extracted from a region
of sky close to M33 (to minimize the difference in the sky X-ray background) and
``filter-wheel closed''  data to produce  a background spectrum.  Using  the SAS
tool  {\it skycast},  a co-added  blank-sky  pn image  was rotated  to the  same
attitude as each observation, and the spatial mask imposed for the bright source
and residual regions. Spectra were then extracted from these areas and scaled to
the same exposure time as the actual observation.  The same process was followed
for  filter-wheel closed  data to  produce a  particle background  spectrum. The
source and blank-sky spectra were then compared in the 8--12 keV band, where the
signal is dominated  by the particle background.  The  difference in this signal
was  compensated for  by addition  of a  relevant fraction  of  the filter-wheel
closed spectrum, the effect of which was to ensure the particle background level
for source and background spectra was the same.


\section{Properties of the galactic X-ray emission}
\label{sec:m33lx}

\subsection{The contribution of luminous point sources}
\label{sec:m33galaxies}

The relative  proximity of M33  means that point  sources in this galaxy  can be
resolved   (and    excluded)   down   to   a   luminosity    threshold   of   $2
\times10^{35}\ergsec$ (0.3--6  keV), which is  $\sim 100$ times deeper  than was
the case  for the majority galaxies  in our earlier study  (OW09). A significant
fraction of X-ray luminosity deriving from HMXBs and LMXBs in M33 can thereby be
excluded. Using a conversion from count-rate to luminosity derived from the 
spectral fit of the bright point source population in the galaxy, we  measure the  
summed X-ray luminosity  (0.3--6 keV) of  the bright
sources contained within the source mask  to be $1.6 \times 10^{39} \ergsec$, to
which the  bright ULX, M33 X-8  makes a dominant contribution.  We determine the
X-ray  luminosity of  this individual  source (averaged  across the  set  of M33
survey  observations)  to  be  $1.2\times10^{39}\ergsec$ (0.3--6  keV),  broadly
consistent   with    the   measurements   reported    by   \citet{grimm05}   and
\citet{misanovic06},  who  report   a  luminosity  of  $8.3\times10^{38}\ergsec$
(0.2--4.5  keV).  This  source  is  detected within  $0.6\arcs$  of the  optical
nucleus  of the galaxy  (\citealt{dubus02}), but  short-term variability  in its
X-ray emission suggests that it is  not the galactic nucleus itself. Analysis by
\citet{foschini04}  suggests   that  the  object   is  a  black  hole   of  mass
$\approx$10\Msun~    accreting     at    a    super-Eddington     rate.    

Table  \ref{table:m33gal_xray}  summarizes  the  distribution across  the  soft,
medium and hard  energy bands of the integrated  luminosity contained within the
bright-source  region (with the  conversion from  count rate  to flux  and hence
luminosity  estimated   from  the  best-fitting  spectral   models  reported  in
\S\ref{sec:m33spec:res}).    The   X-ray    luminosity    pertaining   to    the
residual-emission component  is similarly reported.  The quoted  \lx~figures are
corrected  for foreground  galactic  absorption.  Correction  factors were  also
applied for spillover of source counts into the residual-emission region and for
the underlying extended-emission component contained within the source mask.

It  is possible to  estimate the  integrated X-ray  luminosity of  point sources
below   our   luminosity  threshold,   assuming   such   sources  have   similar
characteristics  to  the resolved  source  population.   For  this, we  use  the
\chandra M33  observations of \citet{grimm05}, who  study a similar  area of the
galaxy  as  considered here.   \cite{grimm05}  derive an  XLF  with  a slope  of
-(0.74--0.78), which  we use to estimate  the total integrated  \lx~ for sources
between $2 \times  10^{35} \ergsec$ and $1 \times  10^{34} \ergsec$. The results
are  given in  Table  \ref{table:m33gal_xray}, and  show  that unresolved  X-ray
binaries do not contribute significantly to the residual X-ray emission observed
in  the soft  band, but  may  form a  significant contribution  to the  emission
observed above 1 keV. This  implies that the residual-emission component largely
arises as the integrated emission of lower-luminosity source populations such as
cataclysmic variables and active binaries plus truly diffuse emission associated
with the  inner disk of  M33, presumably energized  by the collective  effect of
supernovae explosions and stellar winds.


\subsection{Morphology of the soft X-ray emission}
\label{sec:m33morph}

M33 contains a two-armed grand-design spiral structure with spiral arms 
extending to a radius of 10\arcm~as observed in the near infrared 
(\citealt{regan94}), along with a bar structure in the inner 1.5\arcm. 
Far infrared and radio observations confirm this structure, with the southern 
spiral arm the most prominent emission region in the galaxy across all of these wavebands 
(\citealt{hippelein03}; \citealt{tabatabaei07}). The soft X-ray image of M33 
(Fig. \ref{fig:m33xuv1}) clearly shows substantial, apparently diffuse 
emission distributed across the inner disk of the galaxy, with the overlaying pattern of 
the spiral structure particularly pronounced in the region of the southern arm. 
Narrow X-ray structures are similarly observed along the northern spiral arm, 
although they are noticeably less luminous. The soft X-ray emission
can be traced out to a radius of 15\arcm~beyond which the signal drops
below the level of the X-ray background.    

Fig.  \ref{fig:m33xuv1} shows a comparison of the soft (0.3--1 keV) X-ray
morphology  of M33  with corresponding  FUV  ($\lambda_{eff}\approx1528$\AA) and
near-infrared K-band measurements.  The FUV  data are from \galex (\citealt{gil07}) 
and the K-band data from the 2MASS survey (\citealt{skrutskie06}; mosaic images provided by 
T. Jarrett (IPAC)).  The \galex and K-band
images were  resampled to the same spatial  scale as the \xmmn  images, and were
then lightly  smoothed with a gaussian mask  with $\sigma\approx4\arcs$ (roughly
matching the \xmm point-spread function).  The FUV emission traces the locations
of  recent star formation, which  in turn  map out  an underlying
spiral  structure. In  contrast the  K-band emission  shows a  relatively smooth
azimuthal distribution coupled  with a sharp decrease in  the surface brightness
with  increasing galactocentric radius.   The soft  X-ray morphology  appears to
share some of the characteristics of both FUV and K-band images, and the 
association of the X-ray emission with each of these components is explored 
more quantitatively in Section \ref{sec:m33star}.


\section{Spectral Analysis}
\label{sec:m33spectrum}

The  methodology outlined  in \S\ref{sec:m33diff_extr}  was employed  to extract
bright-source and residual-emission spectra  for the three observations identified
in  Table \ref{table:m33obs}. Only  EPIC pn  data were  used in  the spectral
fitting, since the pn camera has  superior sensitivity to the MOS cameras in the
soft  band.    The  bright-source  spectrum  was  fitted   separately  for  each
observation to allow for  possible spectral variations between observations (and
also to  take account of the  fact that some  of the sources encompassed  by one
observation might fall outside the field  of view in another), whereas the three
residual-galaxy  spectra  were  fitted  simultaneously. The
spectral fitting was carried out using the software package XSPEC version 12.5.


\subsection{Spectra of the bright-source region}
\label{sec:m33spec:sources}

\begin{figure*}
\centering
\rotatebox{270}{\scalebox{0.55}{\includegraphics{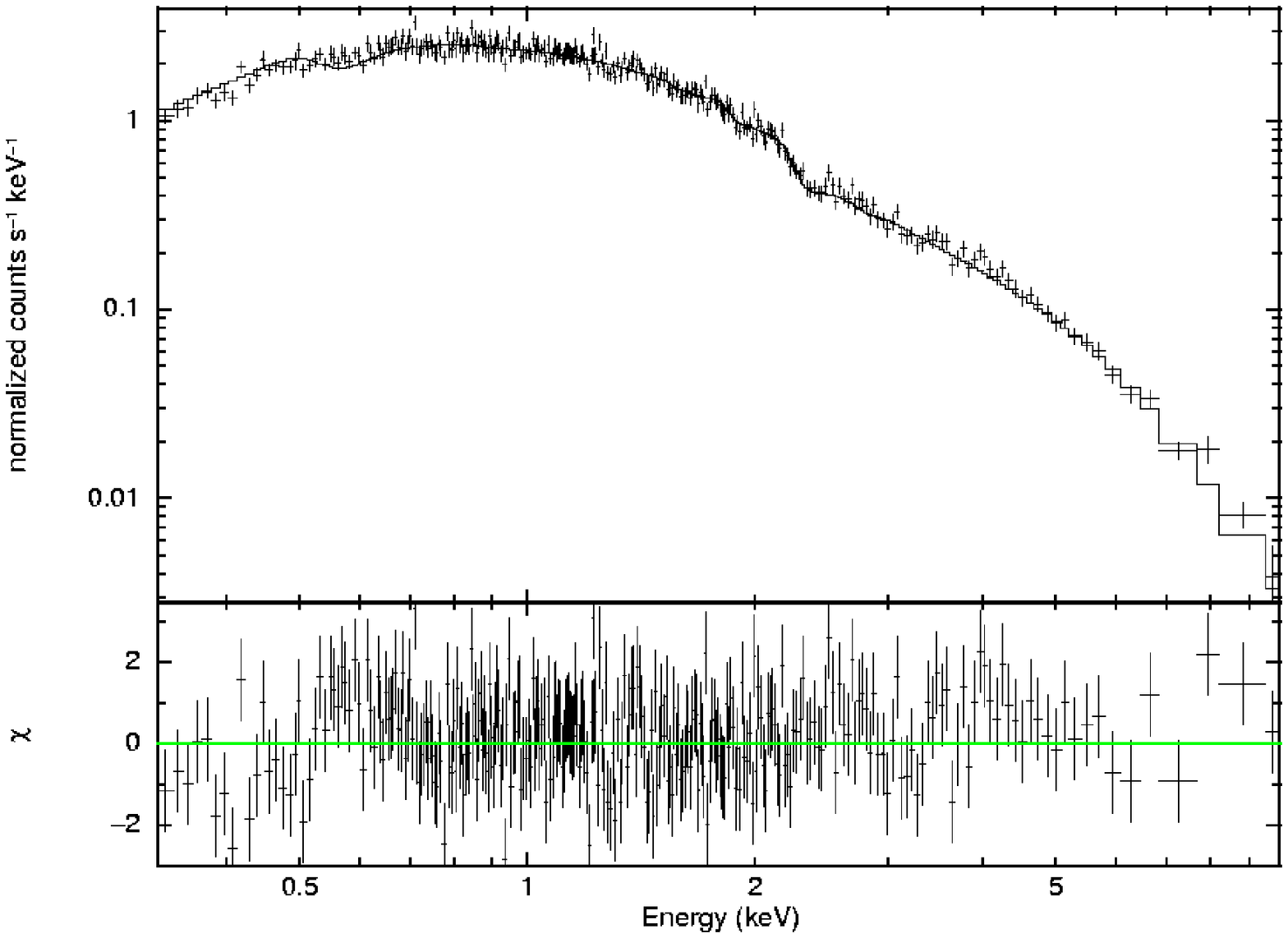}}}
\rotatebox{270}{\scalebox{0.55}{\includegraphics{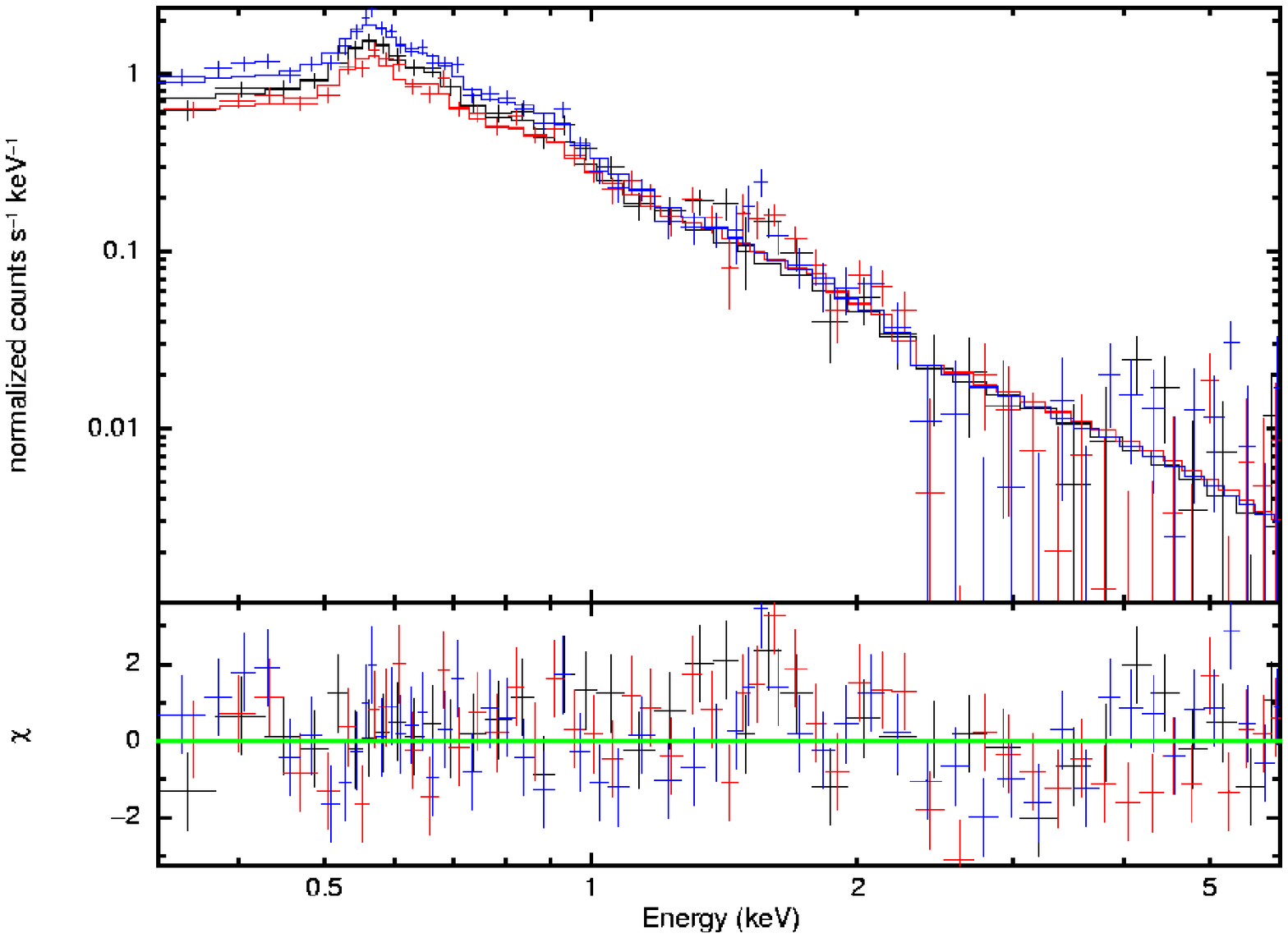}}}
\caption{The EPIC spectra for M33. {\it Top  panel:} 
The spectrum of the bright-source regions from one observation. {\it Bottom panel:} 
The simultaneously fitted spectra of the residual-galaxy region.  In all cases the solid 
line corresponds to the best-fit spectral model (see text). The $\chi^{2}$ residuals 
with respect to the best-fitting model are also shown in both panels.}
\label{fig:m33spec}
\end{figure*}

As noted earlier, the X-ray luminosity of the bright-source population
of M33 is dominated by M33 X-8, the ULX close to the nucleus of the galaxy. 
It follows that our bright-source spectra will also be dominated by 
this source. We conducted the spectral fitting by first extracting
and then modelling the spectrum of the ULX alone using the combination
of a multicolour black-body  disk component and a power-law continuum,
both subject to some intrinsic absorption (see \citealt{foschini04}).
The spectrum  of the ULX appears to change marginally between 
observations,  reflecting the  long-term spectral variability of 
this source.  Next we modelled the bright-source spectra extracted 
from the full masked region by combining an unabsorbed power-law 
component with the ULX contribution, the former representing the
emission of the other bright sources (mainly LMXBs and HMXBs) in M33. 
Fig. \ref{fig:m33spec} ({\it top panel}) illustrates the bright-source 
spectrum and derived composite best-fit spectral form for one of the 
observations. The separate contributions of M33 X-8 and the other 
bright sources to the composite bright-source spectrum are listed in 
Table \ref{table:m33spec_fits} for each observation.

We estimate that $\sim 3\%$ of the flux from the ULX leaks beyond
the source mask into the residual galaxy region, whereas the
corresponding value for the other bright sources is $\sim 9\%$
(the difference reflecting the use of a fixed surface 
brightness cut to define the source mask). These percentages 
were used when modelling the bright-source spillover
into the residual-emission region.


\begin{table*}
\caption{Parameters of the best-fitting models to the spectra of the bright-source and 
residual-galaxy regions.}
\centering
\begin{tabular}{lccccccccc}
\hline

Observation  & Region &  Intrinsic \nh  &  Power-law  &  Disc BB  & Cool MEKAL & Hot MEKAL  &  Goodness  &  Cool:Hot \\
& & $\cm2$  &  Index  &  keV  &  keV  & keV  & of Fit & Flux Ratio \\
& & &  Normalization  &  Normalization  &  Normalization  &  Normalization  &  $\chi^{2}$/dof  & (0.3--2 keV) \\
\hline
0102640101  & Bright Source: & $9.3\times 10^{20}$  &  2.61$\pm$0.04  &  1.16$\pm$0.01 &  -  &  -  &  &  -  \\
& M33 X-8 & &  $1.99 \times 10^{-3}$  &  0.318  &   &    &    &    &    \\
\\
 & Bright Source: & -  &  1.70$\pm$0.06  &  - &  -  &  -  &  -  &  -  \\
& Other & &  $5.48 \times 10^{-4}$  &   &   &    &    &    &    \\
\\
 & Composite Fit& -  & - &  - &  -  &  -  & 676/703  &  -  \\
\\
0102642301 & Bright Source: & $3.5\times 10^{21}$  &  2.90$\pm$0.08  & 1.23$\pm$0.03  &  -  &  -  &  - &  -  \\
& M33 X-8 &  &  $3.06 \times 10^{-3}$  &  0.150  &    &    &    &    \\
\\
 & Bright Source: & -  &  2.04$\pm$0.02  & -  &  -  &  -  &  -  &  -  \\
& Other &  &  $1.90 \times 10^{-3}$  &    &    &    &    &    \\
\\
 & Composite Fit& -  & -  &  - &  -  &  -  & 720/723  &  -  \\
\\
0141980801 & Bright Source: & $8.9\times 10^{20}$  &  2.30$\pm$0.03 & 1.16$\pm$0.02  &  -  &  -  &  -  &  -  \\
& M33 X-8 &  &  $2.83 \times 10^{-3}$  &  0.143  &    &    &     &    \\
\\
& Bright Source: & $6.5\times 10^{20}$  & 1.86$\pm$0.07 & -  &  -  &  -  &  -  &  -  \\
& Other &  &  $5.50 \times 10^{-4}$  &   &    &    &     &    \\
\\
 & Composite Fit& -  & -  &  - &  -  &  -  & 709/672  &  -  \\
\\
\\
Combined  & Residual Galaxy  &  -  &  -  &  -  &  0.17$\pm$0.01   & 0.59$\pm$0.03    &  1337/1266    &  4.5  \\ 
& Solar-abundance &  &  &  &   $8.80 \times 10^{-4}$    &   $1.10 \times 10^{-4}$    &   &   \\
\\
Combined  & Residual Galaxy & -  &  -  &  -  &  0.18$\pm$0.01  &  0.63$\pm$0.07  &  1254/1265  &  4.2  \\
& Free-abundance$^{a}$ & &  &   &  $6.35 \times 10^{-3}$  &  $6.01 \times 10^{-4}$    &  &  \\
\\
\hline

\end{tabular}
\\
$^{a}$ - Residual emission fit at 10$\pm$3\% solar abundance.
\label{table:m33spec_fits}
\end{table*}


\begin{table*}
\caption{Physical properties of the diffuse gas present in each galaxy.}
\centering
\begin{tabular}{lcccccccc}
\hline

Galaxy  &  Radius$^{a}$  &  Component   & Electron Density  &  Thermal Energy  &  Cooling Timescale   \\
& kpc  &  (keV)  &  $10^{-3} \eta^{-1/2}\rm ~cm^{-3}$  &  $10^{54} \eta^{1/2} erg$  &  $10^{8}\eta^{1/2}$ yr \\
\hline
\\
M33 &  3.5  &  0.2  &  4.1  &  4.4  &  17 \\
&  &  0.6  &  1.4  &  4.8  &  61  \\
\\

\hline    
\end{tabular}
\\
$^{a}$ - Assumed radius of a putative shallow halo component (see text)\\
\label{table:m33gas_physics}
\end{table*}


\subsection{Spectra of the residual-emission region}
\label{sec:m33spec:res}

The  three  spectral datasets  which  were  selected  as representative  of  the
residual emission  are shown  in Fig. \ref{fig:m33spec}. As a preliminary step these
datasets were checked for short-term variability at $\sim 0.6$ keV indicative of the 
presence of  \ovii~geocoronal SWCX emission. No evidence for such contamination was found 
and in fact the three observations give a consistent measure of the strength of the 
\ovii line when interpreted as thermal emission from M33 (see below). Nevertheless we 
cannot completely rule out the presence of some residual \ovii emission associated with 
heliospheric SWCX, for which the variability timescale would be relatively long.

The residual-emission spectra were modelled with a combination  of two thermal  
plasma (Mekal) components both subject to line-of-sight absorption in the 
Galactic foreground,  \nh $\sim 7.5 \times 10^{20}  \cm2$ (\citealt{kalberla05}).   
Initially the two  thermal components were constrained  to solar metal abundances.   
A ``cool'' thermal  component at kT
$\approx0.2$ keV, together with a  ``hot'' thermal component at kT $\approx0.6$
keV provided  a reasonable fit to the  spectral data (\rchi=1.05 for  the fit to
all three spectra).   Allowing the metallicity of the  two thermal components to
vary independently  resulted in a constrained  fit only for  the cool component.
With the abundances of the two components  tied, the best fit was obtained for a
metallicity relative  to solar of  10$\pm$3\%. The resulting improvement  in the
$\chi^{2}$  (\rchi=0.99 for  the  spectra set  as  a whole)  was highly significant  as
measured by the F-test. Details of these 
spectral fitting results are summarised in Table \ref{table:m33spec_fits}  and the 
resulting best-fitting model spectrum (with subsolar abundances) is illustrated
in Fig. \ref{fig:m33spec}. 

Several authors (\eg \citealt{strickland98}; W07) have noted  that when attempting  to fit
low-resolution  X-ray spectra  pertaining to  complex  multi-temperature plasmas
with  over-simplified one-  or two-temperature  models, then  a  requirement for
strongly  sub-solar abundances  is often  the  outcome, alas  an erroneous  one.
Notwithstanding this cautionary note, in the present case, the introduction of a
sub-solar abundance does improve the  spectral fit and more specifically removes
a  soft excess  otherwise apparent  in  the residual-galaxy  spectrum below  0.5
keV. This  is evidence, albeit tentative,  that the low-metal  abundance may not
necessarily be a  fitting artifact in this case.   Interestingly the adoption of
sub-solar abundances would also compensate  for the apparent soft X-ray excesses
found previously in M51 and M83 (OW09), although for these sources the signal to
noise ratio was insufficient to give useful individual constraints.  In the case
of M33,  the subsolar metallicity inferred  from the soft  X-ray spectroscopy is
also in line with studies of its \hii regions, for which O/H abundances a factor
of 2-3 below solar have been reported (\citealt{crockett06}).

The   X-ray  luminosity   of   the  residual-emission   component   in  M33   is
$1.2\times10^{38}\ergsec$ (0.3--2  keV).  This  is approximately 10  times lower
than  the value inferred  for the  majority of  the galaxies  studied by  OW09 -
consistent with  M33 being a relatively low-mass system in a satellite orbit around
M31.    The  measurement  of   a  thermal   spectrum  which   may  be
characterised  by  temperatures  of  $\approx  0.2$ and  $\approx  0.6$  keV  is
consistent  with many  previous results for  normal and  starburst galaxies
(\eg   \citealt{ehle98};  \citealt{fraternali02};   \citealt{soria03};  \citealt
{kuntz03}; OW09). In M33 the relative contribution of the two thermal components in the
0.3--2 keV  band is 4.2:1, with  the cool component dominant.   M33 is therefore
spectrally similar  to M74, M101 and NGC3184,  as opposed to M51  and M83, where
there is a more equal weighting between the two thermal components.

We  can derive  the physical  properties  of the  diffuse gas  from the  derived
spectral parameters. If we assume that  the majority of the residual emission we
observe  is  truly  diffuse  and  is  contained within  a  cylindrical  disk  of
major-axis radius  15\arcm (\ie a  linear dimension of 3.5  kpc) and half-width
0.5 kpc,  we can estimate the  electron density, the thermal  energy and cooling
timescale  for each  thermal component.   The  results are  summarised in  Table
\ref{table:m33gas_physics}.  The thermal energy  contained in the two components
is  comparable, implying that  these components  they may  be in  rough pressure
balance.


\section[Spatial analysis]{Spatial Analysis}
\label{sec:m33star}

In an  earlier study  (W07), we found  that when  we compared the  observed soft
X-ray  surface  brightness  distribution  of M101  with  corresponding  datasets
measured in several  optical to ultraviolet wavebands, the  best correlation was
obtained  with the  U-band image.   This was  interpreted in  terms of  an X-ray
signal comprised of  two distinct spatial components, namely  a clumpy thin-disk
component  which  traces  the spiral  arms  of  the  galaxy, and  an  underlying
spatially-smooth  component  which  contributes  significantly  to  the  central
concentration of the soft X-ray emission.  The soft X-ray morphology of M33 (see
\S3) appears to bear  some similarity to that of M101 in  that it is possible to
discern both the spiral arm structure  (which clearly dominates the corresponding
FUV image) and a smoother underlying  component (a  template for which  might 
be provided  by the K-band image).

\begin{figure*}
\centering
\rotatebox{270}{\scalebox{0.42}{\includegraphics{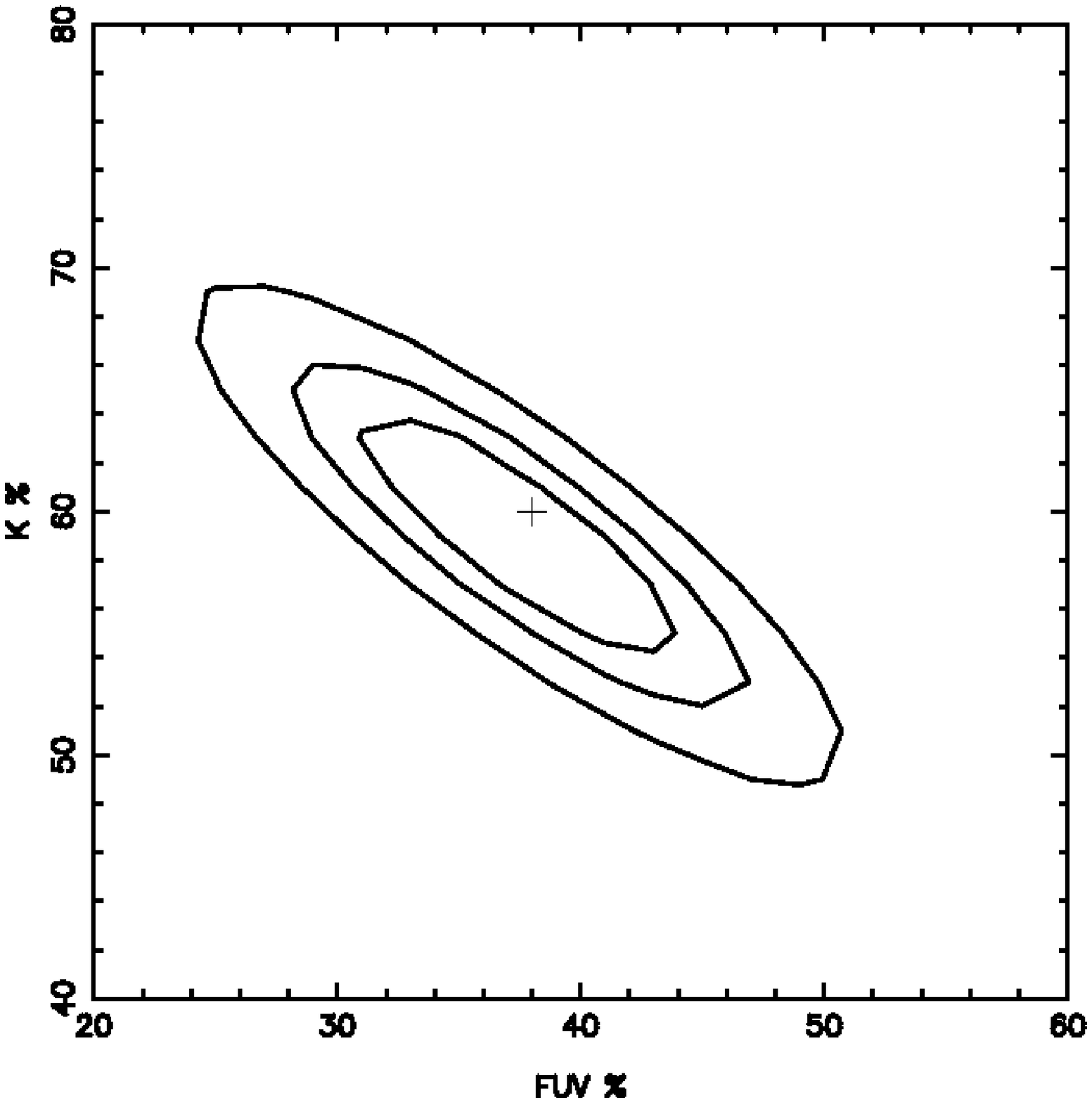}}}
\rotatebox{270}{\scalebox{0.43}{\includegraphics{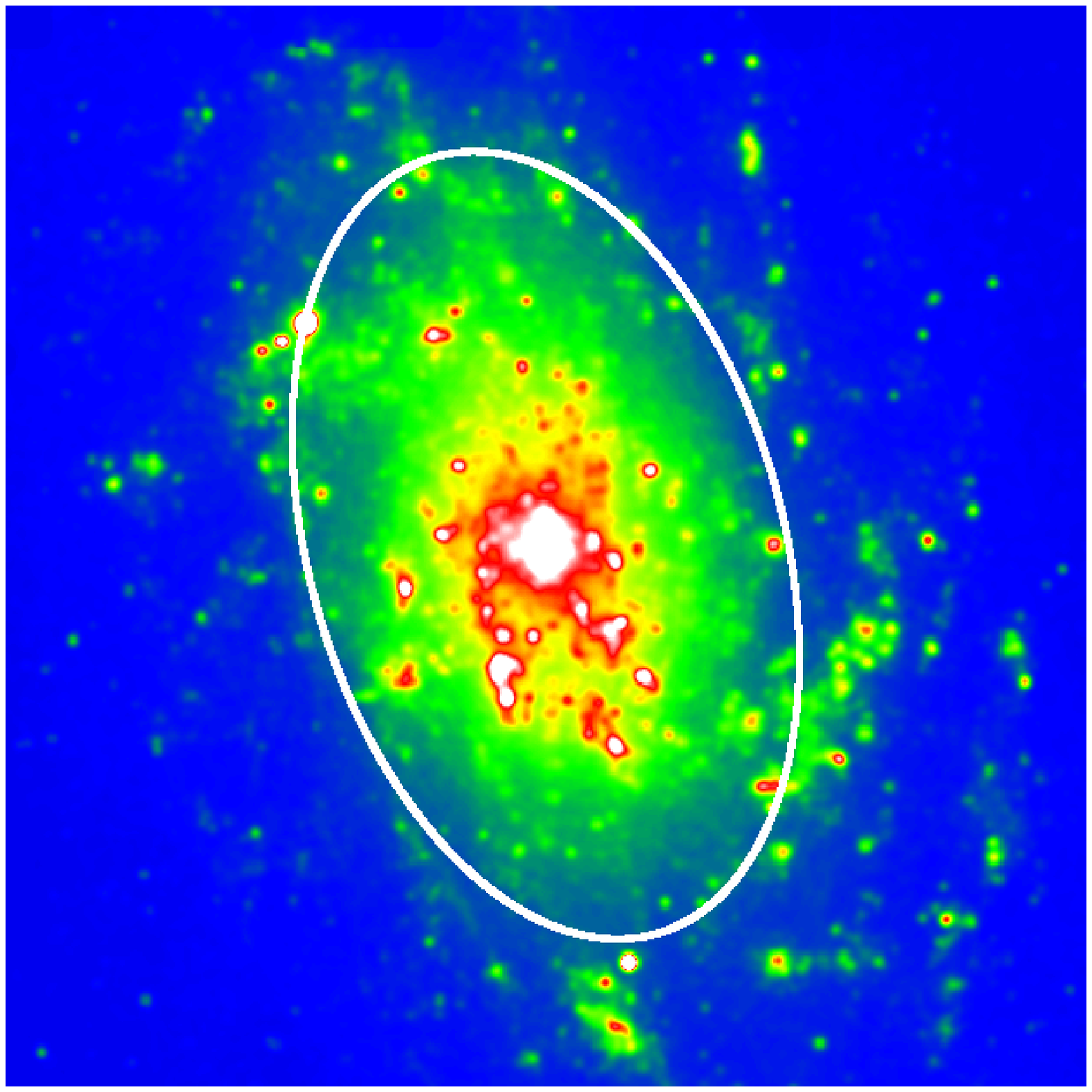}}}
\caption{  
{\it Left-hand panel:}  $\chi^2$ contour plot for the match of the synthetic 
image derived from the FUV and K-band templates to the observed soft X-ray surface 
brightness distribution. The best fit is obtained when the split between the FUV and 
K-band components is approximately 40:60 (marked with the cross). The contours 
represent 68\%, 90\% and 99\% $\chi^2$ confidence intervals. 
{\it Right-hand panel:} 
The synthetic image which best matches the soft X-ray morphology. The displayed image
has  4\arcs~pixels whereas the $\chi^2$ test was performed using 1\arcm~binning (see text).  
The amplitude  scaling  is  logarithmic. The ellipse has  a major-axis 
radius of 15\arcm and corresponds to the X-ray analysis region.}
\label{fig:galaxycontours}
\end{figure*}

To explore  the above idea further, we  have investigated the degree  to which a
``synthetic X-ray image'' produced as a linear combination of the FUV and K-band
images can be made to mimic the  observed soft X-ray emission. To this end, the
soft  X-ray, FUV  and  K-band  images shown  in  Fig.\ref{fig:m33xuv1} were  all
spatially  masked  so as  to  simultaneously  suppress  the bright  X-ray  point
sources, several FUV  regions with very high surface  brightness and some bright
foreground  stars visible in the  K-band image. The  three images were
then compressed to a $1\arcm\times1\arcm$ pixel scale.  A synthetic image formed
by summing scaled-versions of the FUV and K-band images was then compared to the
corresponding soft X-ray  image through the computation of  a $\chi^2$ statistic
(calculated on a  pixel-by-pixel basis and summed over the  set of pixels within
the  elliptical inner  disk region  defined  earlier). By  varying the  relative
contributions  of the  FUV and  K-band  images to  the synthetic  image we  were
thereby able to identify the best-fitting  combination. The best fit was in fact
obtained when  the FUV template image  contributed 40\% of the  total signal (in
the    synthetic    image)    and    the    K-band    equivalently    60\%    (see
Fig.\ref{fig:galaxycontours}).   This  analysis  confirms  a  direct  association
between recent star formation and soft X-ray emission in the inner galactic disk
of  M33.   However,  it  also  points  to a  substantial  contribution  from  an
underlying component correlated with the  K-band light, which in turn traces the
old  stellar  population  of  the  galaxy  and, to  first  order,  the  galactic
stellar-mass distribution (see \S6).


In OW09, we  derived the azimuthally-averaged radial profiles  of the soft X-ray
emission observed in six late-type  spiral galaxies and used these measurements,
in conjunction with published SFR  data, to derive quantitative estimates of the
ratio of the soft X-ray emission per  unit disk area to the local star formation
rate (SFR) per  unit disk area.  Here we determine the  same information for M33
using the  same methodology as  OW09, except that  since M33 is  of intermediate
inclination, with a major/minor axis ratio  of 1.7, in this case it is necessary
to use elliptical  annuli centred on the galactic nucleus  in order to determine
the radial profiles.   Due to the presence of the ULX  near the galactic nucleus
of M33,  we extracted  data from the  spatially-masked soft X-ray  image between
major-axis radii of $2\arcm$ and $14\arcm$, beyond which the soft X-ray emission
falls to near the background level. The corresponding range of linear extent at
the distance of  M33 is 0.46 to 3.2  kpc. In the analysis which  follows, we use
the FUV and K-band radial profiles tabulated in \citet{munoz07}.

\begin{figure*}
\centering
\rotatebox{270}{\scalebox{0.65}{\includegraphics{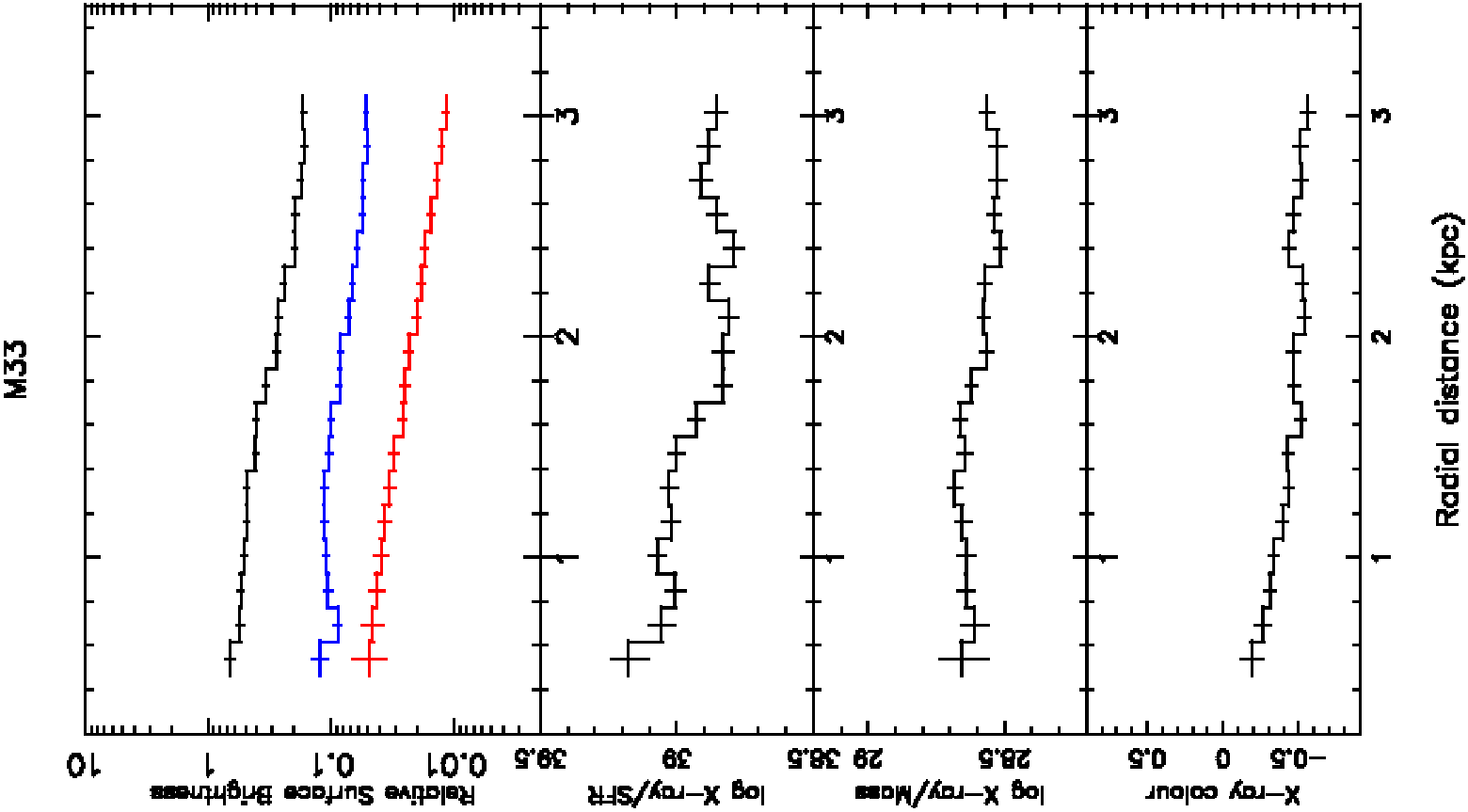}}}
\caption{A comparison of the radial profile of X-ray emission with FUV emission,
SFR and mass distributions in the central disk of M33.  The x-axis refers to the
major-axis radius scaled  to kpc, assuming the distance to M33  is 795 kpc.  The
following  information is  provided: {\it  Top  panel:} The  soft X-ray  surface
brightness versus radius  (upper curve). The radial profile  of the FUV emission
from GALEX  images (middle curve).   The radial profile  of the K-band emission
from 2MASS  images (lower  curve).  {\it  Second panel:} The  ratio of  the soft
X-ray luminosity in $\ergsec$~pc$^{-2}$ (0.3-1 keV) to the local SFR in units of
\Msun~yr$^{-1}$~pc$^{-2}$.   {\it Third  panel:}  The ratio  of  the soft  X-ray
luminosity  in $\ergsec$~pc$^{-2}$  (0.3-1 keV)  to the  K-band derived  mass in
units  of \Msun~pc$^{-2}$.   {\it  Bottom panel:}  Variation  in X-ray  spectral
hardness, (H-S)/(H+S),  versus radius, where H  refers to the  0.8--1.2 keV band
and S to the 0.3--0.8 keV band.}
\label{fig:m33radial_plots}
\end{figure*}

The K-band  and soft  X-ray radial profiles  (Fig.\ref{fig:m33radial_plots}, top
panel)  show  very   similar  rates  of  fall-off  between  0.5   and  3  kpc;
\citet{munoz07} quote a scale length of 1.48 kpc for the K-band photometry. 
In contrast, the FUV  radial profile although of  comparable slope to
the X-ray  and K-band curves  outside of 2 kpc, appears to  flatten at
smaller radii. Although, given the background uncertainties, we are unable to place
quantitative limits on  the level of the soft X-ray emission  much beyond 3 kpc,
visual  inspection  of the  X-ray  images suggest  that  the  soft X-ray  signal
declines very sharply  outside of the inner-disk region.   This rapid switch-off
is reminiscent of a similar effect evident in NGC300, M74, M51 and M101 (OW09).


Fig. \ref{fig:m33radial_plots},  (second panel) shows  a comparison of  the soft
X-ray luminosity in  the 0.3-1 keV band  (per unit disk area) with  the SFR (per
unit  disk   area)  (the  latter  as  tabulated   in  \citealt{munoz07}).   More
specifically the ratio  of these two quantities has a  value of $\approx1 \times
10^{39} \ergsec$  (\Msun~yr$^{-1}$)$^{-1}$, at a  radius 0.5 kpc,  with evidence
for a  modest decline (by a  factor dex$\approx$0.2) across  the sampled region.
In order to  make a direct comparison with the results  for the galaxies studied
in OW09, we need  to apply two corrections as follows: (i)  we convert the X-ray
luminosities to a  broader 0.3-- 2 keV band using  the spectral models discussed
in (\S\ref{sec:m33spec:res} - this results  in an upward scaling of the X-ray/SFR
ratio by 5\%);  (ii) we correct the X-ray luminosities to  a common point source
exclusion threshold of \lx = $1 \times 10^{37} \ergsec$ (0.3-6 keV) (in the case
of M33 this involves integrating  the luminosity encompassed by its point-source
XLF over the range \lx = $2-100 \times 10^{35} \ergsec$) - this results in a further upward
scaling of 50\%. The resulting  estimate of the X-ray/SFR ratio of $1-1.5
\times 10^{39}  \ergsec$ (\Msun~yr$^{-1}$)$^{-1}$ lies towards the  upper end of
the range observed  to date, at a  very similar level to that  pertaining to the
inner disks of M51 and M83
(\cf Fig.5, OW09, where for M33 the parameter log(SFR density)
ranges from -8 to -7.5)\footnote{In the discussion section of OW09, the upper bound of the X-ray to 
SFR ratio was quoted as $5 \times 10^{39}  \ergsec$ (\Msun~yr$^{-1}$)$^{-1}$. This included 
an additional scaling factor of 3.4 arising from the conversion from the SFR mass 
range employed by \citet{munoz07} 
(0.1--100 \Msun) to that used in \citet{mas08} (2--120 \Msun).}.





The soft X-ray radial profile and the stellar-mass radial profile estimated from
the  K-band photometry can  similarly be  compared.  \citet{munoz07}  use K-band
data to evaluate the stellar-mass  radial profile of M33. 
We find that the  soft X-ray/stellar-mass ratio remains relatively constant at
$\approx 4 \times 10^{28} \ergsec\Msun^{-1}$, across the inner disk of M33 (Fig.
\ref{fig:m33radial_plots},   third   panel).    Recent  studies   conducted   by
\citet{revnivtsev07a} and  \citet{revnivtsev07b} have explored  the relationship
between  unresolved X-ray emission  and the  stellar-mass distributions  for the
Galactic ridge  and M32, in an attempt  to derive the X-ray  emissivity per unit
stellar mass in systems with low  SFR. Most likely this emissivity originates in
point  sources associated with  (intermediate) disk  population objects  such as
active binaries  and cataclysmic variables  with \lx ~between  $10^{30} \ergsec$
and $10^{34}  \ergsec$, together with  a contribution from fainter  sources, the
bulk of which will be coronally  active stars. From these and other studies (\eg
NGC3379  and  M31  - \citet{revnivtsev08};  \citet{bogdan08}),  an  X-ray  to
stellar mass ratio  for quiescent systems (which do  not contain substantial hot
gas)  can be  estimated  to  be $4-8  \times  10^{27} \ergsec\Msun^{-1}$  (0.5-2
keV). This is a factor of 5-10 lower than we derive above for M33, implying that
$\approx  10-20\%$  of the  observed  unresolved  X-ray  luminosity in  M33  may
be associated with the  old stellar population. This  level is consistent
with that predicted from the extended XLF derived for LMXB, relatively quiescent 
LMXB, cataclysmic variables and other source types associated with the old
population. 

As  a final investigation  we also  considered the  variation in  X-ray spectral
hardness as deduced  from images produced in two  sub-bands, namely the 0.3--0.8
keV and 0.8--1.2 keV bands. The plot of spectral hardness versus radius shown in
(Fig. \ref{fig:m33radial_plots},  bottom panel)  in essence tracks  the relative
contribution of  the two thermal  spectral components identified earlier  in the
spectral  fitting  (\S\ref{sec:m33spec:res}).   There  is  a  hint  of  spectral
softening  with increasing  galactocentric  radius, suggesting  that the  hotter
thermal component makes  a greater contribution to the  total luminosity towards
the centre of the galaxy.


\section{Discussion}\label{sec:m33disc}

The  strong correlation  observed between  soft X-ray  and FUV  emission  in M33
confirms  a direct linkage  between star  formation and  the production  of soft
X-rays in  this galaxy.  According  to current models  (\eg \citealt{cervino02}:
\citealt{mas08}),  following a burst  of star  formation which  gives rise  to a
massive star  cluster, the FUV emission may  be expected to peak  on a timescale
matching the lifetime of the  most massive stars ($\sim 1-3\times 10^{6}$ years),
whereas the  X-ray signal reaches  a maximum sometime later.  In this scenario,
diffuse  soft X-rays  are produced  through the  heating of  bubbles  within the
interstellar  medium  to  temperatures  of  $10^{6-7}$  K as  a  result  of  the
mechanical energy input from the winds of massive stars created in the starburst
and the eventual  destruction of such stars in  supernovae.  Individual SNR also
contribute,   primarily  during   the free-expansion and adiabatic   phases  of   their  evolution.
Occasionally luminous high-mass X-ray binaries might be formed in such a cluster,
although here we assume that such  sources would be bright enough to be excluded
as resolved point-sources. \citealt{mas08} find  that following the onset of the
starburst the diffuse soft X-ray luminosity increases rapidly over the first few
Myr.   For an  instanteous  burst, the  X-ray  luminosity may  then plateau  and
eventually decline whereas, if the  starburst activity is ongoing then the X-ray
luminosity  may  gradually rise  for  up  to $\sim  30$  Myr  until the  stellar
formation and  destruction process  reach an equilibrium.   Using the  models of
\citet{mas08}, we find  that the X-ray/SFR ratio measured for  the inner disk of
M33 matches  the predictions at  $\sim 10$ Myr  (after the onset of  an extended
burst  of star  formation activity),  assuming the  efficiency of  conversion of
mechanical energy from supernovae into X-rays is 1\% (\cf OW09).  Studies of the
star  formation  history of  M33  (\citealt{wilson88}; \citealt{wilson95})  have
shown that several strongly emitting  \hii regions have undergone bursts of star
formation  on the  timescale  of 10  Myr,  which indicates  that  this model  is
realistic.

Through a quantitative comparison of  the observed soft X-ray surface brightness
distribution with a  synthetic image, we found that roughly  $40\%$ of the X-ray
signal  had a spatial  distribution similar  to that  seen in  the FUV  with the
remaining  $60\%$  better  matching  the  K-band distribution.  In  effect  this
represents a 40:60  split between a clumpy spiral-arm  distribution and a smooth
disk  distribution.  However, in  \S6 we  also discussed the  fact that
perhaps $20\%$ of the total observed soft X-ray emission might be {\it directly}
associated with the  old stellar population of the galaxy  (for which the K-band
light again serves  as a tracer).   If we  subtract this fraction  we are left  with a
roughly  equal split between the two  inferred spatial components  of the  soft X-ray
emission. 

We are  now in a  position to  add some refinements  to our earlier  analysis in
which  we modelled  the X-ray  emitting  gas as  filling a  cylindrical disk  of
half-thickness  0.5 pc, extending  to a  galactocentric radius  of 3.5  kpc.  As
the spiral arm regions occupy at most $\sim20\%$ of the inner galactic disk by area,
we may model 40\% of the emission (directly connected to the spiral arm regions)
as  filling 20\%  of the  disk with  a  thickness of,  say, 200  pc.  With  this
adjustment, the  cooling timescale of the gas  is $\sim 2\times10^{8}\eta^{1/2}$
yr, where  $\eta$ is the filling  factor of the  gas.  On the basis of the dynamical 
analysis  of M33 performed by \citet{puerari93}, narrow features initially distributed 
in a spiral arm pattern would  be  completely  smeared   out  by  differential  
galactic  rotation on timescales of $2  - 3\times10^{8}$ yr. For such narrow features
to survive we need a much shorter cooling timescale (assuming radiative cooling is the
dominant process) implying a filling factor much less than unity. 
Setting  $\eta \sim 10^{-3}$  gives a cooling timescale  of $6 \times 10^{6}$
years, comparable to  the inferred  lifetime of  hot bubbles  observed in  our own
Galaxy  (\citealt{egger95}; \citealt{breitschwerdt09}).  As noted earlier, this is 
also roughly the timescale on which the soft X-ray production maximises following 
the onset of a starformation episode.  Given  the rotational period and pattern speed
of  the galaxy  (\citealt{puerari93}), this  corresponds to  a $\sim 6\dg$ azimuthal
offset at a  galactocentric radius of 2  kpc, between the
X-ray emission and the FUV emission (with the former leading the latter since at
2  kpc we are  within the  corotational radius).   Such an  offset would  not be
observable  in M33  in \xmmn  data and,  in  fact, such effects have  yet to  be
identified  even in the  higher spatial  resolution \chandra  observations of other
spiral galaxies (\citealt{tyler04}).

Even after allowing for the contribution of the old stellar population, 
we are left with a substantial fraction of the soft X-ray 
emission originating in a smoothly distributed component.  
Since M33 has an intermediate inclination, it
is not possible to  disentangle disk emission from a lower-halo component.   
However, \citet{strickland00}, in a  study of 9  nearby edge-on
spiral  starburst and  normal galaxies,  have managed to distinguish between 
unresolved disk and lower-halo components and find that the former 
dominates across their sample. In all but the
systems with  highest SFR,  the luminosity of  the halo  emission is at  least a
factor of 3-4 lower than that observed in the disk. In the case of M33 it therefore 
seems reasonable to assume that the bulk of the smoothly-varying component 
is confined to the disk but located both in the spiral arms and inter-arm regions. 
A cooling timescale of $\sim 10^{9}\eta^{1/2}$ yr for this component is consistent
with its smooth distribution across the inner disk of M33.


\section{Conclusions}
\label{sec:m33conc}

We have used archival \xmmn  observations to examine the residual X-ray emission
observed from M33 after the exclusion of the bright point source population to a limit of \lx
$~> 2  \times 10^{35} \ergsec$.  Using the  same methodology as in  OW09, we have
investigated the spectral and spatial  properties of the X-ray emission within an inner
disk region extending up to 3.5 kpc from the nucleus of the galaxy.

The observed X-ray spectrum can be modelled as thermal 
emission with cool and hot components of $\approx  0.2$ keV  and
$\approx  0.6$ keV respectively, with the cooler emission providing the dominant contribution
to the luminosity. There is some evidence for a subsolar metallicity 
consistent with other indicators of low-metallicity in M33. 

The  strong correlation established between  X-ray and  FUV  morphologies
confirms the close linkage between X-ray emission and recent star formation.
Detailed comparison of  soft X-ray and FUV radial profiles in  the inner disk of
M33 reveals  the ratio of X-ray emission  to SFR to be  $1-1.5 \times 10^{39}
\ergsec$ (\Msun~yr$^{-1}$)$^{-1}$ (in  the 0.3-2 keV band).  This 
matches the  predictions of \citet{mas08} for an extended
burst of star  formation occurring 10 Myr ago, with  an efficiency of mechanical
energy conversion to X-rays of $\sim1\%$.  The soft X-ray emission
to mass ratio found for  M33 is $4 \times 10^{28}
\ergsec\Msun^{-1}$, a  factor of  5 higher than  the corresponding  value for dwarf
elliptical galaxies and spirals with low SFR. This implies that up to $\sim20\%$ 
of the observed soft X-ray emission originates in the old stellar source population, 
in source types such as cataclysmic variables and active binaries. 

With the contribution of the old stellar population subtracted, the soft X-ray emission 
is found to be equally split between two spatial components, one which closely traces 
the spiral arms of the galaxy and the other more smoothly distributed across the inner 
disk region. The constraints on the cooling timescale implied by the presence of spiral 
features in the soft X-ray images suggest the presence of a highly clumped component, 
encompassing sites of on-going starformation, \hii regions and hot gas bubbles. The nature of the
smoothly distributed component is much less certain. Plausibly it may represent
the integrated emission from a whole range of
processes and source types including supernovae occurring in the interarm regions, individual
sources and source complexes with luminosity not far below the applied
luminosity threshold and an accumulated distribution of hot gas 
which has managed to leak away from the site of its original production.



\section*{Acknowledgments}

RAO acknowledges the receipt of a 
PPARC/STFC research studentship. We thank Tom Jarrett (IPAC) for providing 
us with 2MASS mosaic images and advice, and Jenny Carter for assisting us with 
detection of SWCX in our observations. We thank the referee, Steve Snowden, for comments 
and suggestions which have helped us to improve this paper.

\label{lastpage}

\end{document}